\documentclass[pre, preprint, eqsecnum, showpacs, epsfig]{revtex4-1}
\usepackage{array,longtable}
\usepackage{graphics}
\usepackage{graphicx}
\usepackage[colorlinks=true,linkcolor=blue,urlcolor=blue,citecolor=blue]{hyperref}
\usepackage{amsmath}

\def \vG {$\vec{G}$ }
\def \vR {$\vec{R}$ }

\def \rhor{{$ \rho(\vec{r}) $ }}
\def \rhol{$ \rho_{l} $ }
\def \rhos{$ \rho_{s} $ }
\def \rhoG{$ \rho_{G} $ }
\def \muG{$ \mu_{G} $ }

\def \mc {\multicolumn}

\begin{document}
\title{Fluid - solid transition in simple systems using density 
functional theory}
\author{Atul S. Bharadwaj and Yashwant Singh}
\affiliation{Department of Physics, Banaras Hindu University, 
Varanasi-221 005, India.}
\date{\today}

\begin{abstract}
 A free energy functional for a crystal proposed by Singh and Singh (Europhysics Letters \textbf{88}, 16005 (2009)) which contains both the symmetry-conserved and symmetry-broken parts  of the direct pair correlation function has been used to investigate the fluid-solid transition in systems interacting via purely repulsive WCA Lennard - Jones (RLJ) potential and the full Lennard - Jones (LJ) potential. The results found for freezing parameters for the fluid - face centred cubic (fcc) crystal transition are in very good agreement with simulation results. It is shown that although the contribution made by the  symmetry broken part to the grand thermodynamic potential at the  freezing point is small compared to that of the symmetry conserving part, its role is crucial in stabilizing the crystalline structure and on values of freezing parameters. The effect of attractive part of the LJ potential on the freezing parameters is found to be small, confirming the view that the fluid - solid transition is primarily determined by the repulsive part of the potential.
\end{abstract}


\maketitle

\section{Introduction}
 The fluid-solid transition in three dimensions is a first order phase  transition in which continuous symmetry of the fluid is broken into one  of the Bravais lattices. The density functional theory (DFT) of  freezing, first proposed in 1979 by Ramakrishnan and Youssuf (RY) \cite{ramakrishnan-PRB-1979} has extensively been used to study this  transition. The central quantity in this theory is the reduced  Helmholtz free energy of both the crystal, $ A[\rho] $, and the fluid,  $ A(\rho_{l}) $ \cite{haymet-JCP-1981}. For a crystal, $ A[\rho] $ is a  unique functional of single particle density distribution \rhor whereas  for the fluid  $ A(\rho_{l}) $ is simply a function of fluid density $  \rho_{l} (=N/V $, \textit{N} being the number of particles in volume $  V) $. The density functional formalism is used to find expression for $  A[\rho] $ (or for grand thermodynamic potential) in terms of \rhor and  the direct pair correlation function (DPCF). Minimisation of this  expression with respect to \rhor leads to an expression that relates  \rhor to the DPCF \cite{ysingh-PHYSREP-1991}. The DPCF that appears in  these expressions corresponds to crystal and is functional of \rhor.  When this functional dependence is ignored by replacing the DPCF by  that of the coexisting uniform fluid \cite{ramakrishnan-PRB-1979} or by  that of an ''effective uniform fluid" \cite{tarazona-PRA-1985,curtin-PRA-1985}, the free energy functional becomes approximate and fails to  provide an accurate description of freezing transition for a large  class of intermolecular potentials \cite{kuijper-JCP-1990,wang-JCP-1999}.

 A free energy functional in which the functional dependence of DPCF on  \rhor has been taken into account has recently been proposed  \cite{mishrap,*mishrap-JCP-2007,singhsl-EPL-2009} and applied to steady  freezing of fluids in two- and three-dimensions. The results found for  the isotropic-nematic transition \cite{mishrap,*mishrap-JCP-2007},  fluid-solid transition in systems interacting via the inverse power  potential $ u(r)=\epsilon\left( \sigma/r \right)^{n} $ where $  \epsilon, \sigma $ and \textit{n} are potential parameters and  \textit{r} is molecular separation \cite{singhsl-EPL-2009,jaiswala-PRE-2013,bharadwajas-PRE-2013} and freezing of fluids of hard spheres  into crystalline and glossy phases\cite{singhsl-PRE-2011} are very  encouraging. Furthermore, the theory predicts that the fluids  interacting via the inverse power potentials freeze into a face- centred-cubic (fcc) lattice when the potential parameter $ n\geq 6.5 $  and into the body-centred-cubic (bcc) lattice when $ n\leq 6 $ and the  fluid-bcc-fcc triple point is at $ 1/n = 0.158 $\cite{bharadwajas-PRE-2013}. These results are in very good agreement with simulation  results. To best of our knowledge this is the only free energy  functional which correctly describes the relative stability of the two  cubic phases.

 In this paper we apply the theory to investigate freezing of fluids  interacting via the 6-12 Lennard-Jones(LJ) potential,

\begin{align}
u(r) = 4\epsilon\left( \left(\frac{\sigma}{r}\right)^{12} - \left(\frac{\sigma}{r}\right)^{6}\right),               \label{1.1}
\end{align} 
 where $ \epsilon $ and $ \sigma $ are potential parameters, and compare  our results with the results found from other free energy functionals as well as with simulation results. Also, in order to estimate the role played by the attractive and repulsive parts of the LJ potential in formation of crystalline structure at the freezing point we  consider the purely repulsive Weeks-Chandler-Anderson (WCA) reference  potential defined as\cite{WCAS}

\begin{align}
 u_{R}(r)=\begin{cases}
     u(r)+\epsilon   & \qquad \text{for} \qquad r \leq r_{m} \\
     0        & \qquad \text{for} \qquad r \geq r_{m},  
     \end{cases}  
\label{1.2}
\end{align} 

 where $ r_{m} (= 2^{1/6} \sigma ) $ is the value of r at which the LJ potential has its minimum value. Henceforth, we refer this potential as a reference Lennard-Jones (RLJ) potential. While the LJ potential mimics characteristics of interaction potential of the rare-gas elements and even of some molecular systems, the RLJ potential is used to model interactions in polymers \cite{polymer-1} and dendrimers \cite{dendrimer-1,*dendrimer-2}. The freezing parameters for these systems calculated by de Kuijper et al \cite{kuijper-JCP-1990} using RY free energy functional (RY-DFT), the modified weighted density approximation (MWDA) \cite{denton-PRA-1989} and the modified effective liquid approximation (MELA)\cite{BAUS-MELA} show that these theories fail to give satisfactory description  of the transition.

 The paper is organized as follows: In Sec II we give a brief description of the free -energy functional for a crystal that contains both the symmetry conserving and the symmetry broken parts of DPCF. In Sec III we describe calculation of these functions and report results. In Sec IV the freezing parameters are calculated and compared with simulation results as well as with results found from other (approximate) theories. The paper ends with a brief summary and conclusions given in Sec V.

\section{Theory}

 The formation of a crystalline structure defined by a set of discrete vectors \vR  at the freezing point leads to emergence of a qualitatively new contribution in distribution of particles \cite{singhsl-EPL-2009,singhsl-PRE-2011,jaiswala-PRE-2013,bharadwajas-PRE-2013}. The correlation functions in a crystal can therefore be written as a sum of two qualitatively different contributions; one that preserves the continuous symmetry of the fluid and one that breaks it and vanishes in the fluid \cite{bharadwajas-PRE-2013}. Thus for the DPCF in a crystal we write 

\begin{align}
c(\vec{r_{1}},\vec{r_{2}}) = c^{(0)}(\mid\vec{r_{2}}-\vec{r_{2}}\mid; \rho_{s}) + c^{(b)}(\vec{r_{1}},\vec{r_{2}};[\rho]),  \label{2.1}
\end{align}

 where $ c^{(0)} $ and $ c^{(b)} $ represent respectively, the symmetry conserving and symmetry broken contributions. Note that $ c^{(0)} $ depends on the magnitude of inter-particle separation \textit{r} and is a function of average crystal density, $ \rho_{s} $ while $ c^{(b)} $ is functional of \rhor (indicated by square bracket) depends on position vectors $ \vec{r_{1}} $ and $ \vec{r_{2}} $ and is invariant only under a discrete set of translations corresponding to lattice vectors \vR. The DPCF $ c(\vec{r_{1}},\vec{r_{2}}) $ is related with the total correlation function $ h(\vec{r_{1}},\vec{r_{2}}) $ through the Ornstien - Zernike (OZ) equation \cite{Hansen-book}. The reduced free energy functional $ A[\rho] $ has an ideal gas part,

\begin{align}
A_{id}[\rho] = \int d\vec{r} \rho(\vec{r}) \left(ln\rho(\vec{r})\Lambda)-1\right),   \label{2.2}
\end{align}

 where $ \Lambda $ is cube of thermal wavelength associated with a particle, and the excess part $ A_{ex}[\rho] $ arising due to interparticle interactions. This excess part $ A_{ex}[\rho] $ is related to $ c(\vec{r_{1}},\vec{r_{2}}) $ as \cite{haymet-JCP-1981,BAUS-MELA},

\begin{align}
\dfrac{\delta^{2}A_{ex}[\rho]}{\delta\rho(r_{1}) \delta\rho(r_{2})} = -c(\vec{r_{1}},\vec{r_{2}}).   \label{2.3}
\end{align}

Using Eq(\ref{2.1}) one can rewrite Eq(\ref{2.3}) as

\begin{align}
\dfrac{\delta^{2}A_{ex}^{(0)}[\rho]}{\delta\rho(r_{1}) \delta\rho(r_{2})} = -c^{(0)}(\mid\vec{r_{2}}-\vec{r_{1}}\mid;\rho_{s}),   \label{2.4}
\end{align}

\begin{align}
\dfrac{\delta^{2}A_{ex}^{(b)}[\rho]}{\delta\rho(r_{1}) \delta\rho(r_{2})} = -c^{(b)}(\vec{r_{1}},\vec{r_{2}};[\rho]),   \label{2.5}
\end{align}

where $ A_{ex}^{(0)}[\rho] + A_{ex}^{(b)}[\rho] = A_{ex}[\rho] $.

 The expressions for $ A_{ex}^{(0)} $ and $ A_{ex}^{(b)} $ are found from functional integrations of Eqs(\ref{2.4}) and (\ref{2.5}), respectively. In this integration the system is taken from some initial density to the final density distribution along a path in the density space, the result is independent of the path of integration.  These integrations give \cite{jaiswala-PRE-2013, bharadwajas-PRE-2013},

\begin{align}
A_{ex}^{(0)}[\rho] = A_{ex}(\rho_{l}) + \beta \mu -ln\left(\rho_{l}\Lambda\right) - \frac{1}{2}\int d\vec{r_{1}} \int d\vec{r_{2}} \left(\rho(\vec{r_{1}})-\rho_{l}\right) \left(\rho(\vec{r_{2}})-\rho_{l}\right) {\bar{c}}^{(0)}(\mid\vec{r_{2}}-\vec{r_{1}}\mid;\rho_{l}),        \label{2.6}
\end{align}

and 

\begin{align}
A_{ex}^{(b)}[\rho] = - \frac{1}{2}\int d\vec{r_{1}} \int d\vec{r_{2}} \left(\rho(\vec{r_{1}})-\rho_{s}\right) \left(\rho(\vec{r_{2}})-\rho_{s}\right) {\bar{c}}^{(b)}(\vec{r_{1}},\vec{r_{2}}),        \label{2.7}
\end{align}

where

\begin{align}
{\bar{c}}^{(0)}(\mid\vec{r_{2}}-\vec{r_{1}}\mid;\rho_{l}) = 2 \int_{0}^{1} d\lambda \lambda \int_{0}^{1}d\lambda' {c}^{(0)}\left(\mid\vec{r_{2}}-\vec{r_{1}}\mid;\rho_{l} + \lambda \lambda'(\rho_{s}-\rho_{l}) \right),        \label{2.8}
\end{align}

\begin{align}
{\bar{c}}^{(b)}(\vec{r_{1}},\vec{r_{2}}) = 4 \int_{0}^{1} d\xi \xi \int_{0}^{1}d\xi'\int_{0}^{1} d\lambda \lambda \int_{0}^{1}d\lambda' {c}^{(b)}(\vec{r_{1}},\vec{r_{2}}; \lambda \lambda' \rho_{s}; \xi \xi' \rho_{G}).         \label{2.9}
\end{align}

 In above equations, $ A_{ex}(\rho_{l}) $ is reduced excess free energy of the  coexisting fluid of density \rhol and chemical potential $ \mu $, $ \rho_{s} = \rho_{l}(1+\Delta\rho^{*}) $ is average density of the crystal, $ \beta = (k_{B} T)^{-1} $ is the inverse temperature in unit of the Boltzmann constant $ k_{B} $. The order parameter \rhoG which appears in the expansion of \rhor in the Fourier series as,

\begin{align}
\rho(\vec{r}) = \rho_{s} + \sum_{G \neq 0} \rho_{G} e^{i\vec{G} \cdot \vec{r}},  
 \label{2.10}
\end{align}
 is amplitude of density wave of wavelength equal to ${{2\pi}/{\mid\vec{G}\mid}} $ where \vG is reciprocal lattice vector (RLV). The summation in Eq(\ref{2.10}) is over the complete set of RLV of a given crystal.

 The free energy functional $ A[\rho] $ for a crystal is sum of $ A_{id} $, $ A_{ex}^{(0)} $ and $ A_{ex}^{(b)} $. Thus

\begin{align}
A[\rho] &=\int d\vec{r} \rho(\vec{r})\left(ln(\rho(\vec{r})\Lambda)-1\right)+ A_{ex}(\rho_{l}) + \left[ \beta \mu -ln\left(\rho_{l}\Lambda\right)\right] \int d\vec{r} \left(\rho (\vec{r}) - \rho_{l} \right)   \nonumber  \\
&\qquad - \frac{1}{2}\int d\vec{r_{1}} \int d\vec{r_{2}} \left(\rho(\vec{r_{1}})-\rho_{l}\right) \left(\rho(\vec{r_{2}})-\rho_{l}\right) {\bar{c}}^{(0)}(\mid\vec{r_{2}}-\vec{r_{1}}\mid;\rho_{l})  \nonumber \\
&\qquad - \frac{1}{2}\int d\vec{r_{1}} \int d\vec{r_{2}} \left(\rho(\vec{r_{1}})-\rho_{s}\right) \left(\rho(\vec{r_{2}})-\rho_{s}\right) {\bar{c}}^{(b)}(\vec{r_{1}},\vec{r_{2}}).    \label{2.11} 
\end{align}

 This expression of $ A[\rho] $ which includes both the symmetry conserving and symmetry broken contributions of the DPCF is exact; no approximation has been used in deriving it. In the RY free energy functional the contribution arising due to $ c^{(b)} $ was neglected.

 In locating the freezing transition, the grand thermodynamic potential defined as

\begin{align}
 - W = A - \beta \mu \int d\vec{r} \rho(\vec{r})           \label{2.12}
\end{align}

 is generally used as it ensures that the pressure and the chemical potential $ \mu $ of the two phases remain equal at the transition. The fluid-solid coexistence is obtained when $ \Delta W = W_{l} - W = 0 $, where $ W_{l} $ is the grand thermodynamic potential of the coexisting fluid, and $ \delta W/\delta \rho(r) = 0$ are simultaneously satisfied.

 The expression for $ \Delta W $ is found to be \cite{jaiswala-PRE-2013, bharadwajas-PRE-2013}

\begin{align}
\Delta W &= \int d\vec{r} \left[\rho(\vec{r}) ln\left(\frac{\rho(\vec{r})}{\rho_{l}}\right) - \left(\rho(\vec{r}) - \rho_{l}\right) \right] \nonumber \\
&\qquad - \frac{1}{2} \int d\vec{r_{1}} \int d\vec{r_{2}} \left(\rho(\vec{r_{1}}) - \rho_{l}\right) \left(\rho(\vec{r_{2}}) - \rho_{l}\right) {\bar{c}}^{(0)}\left(\mid \vec{r_{2}} - \vec{r_{1}} \mid \right) \nonumber \\
&\qquad - \frac{1}{2} \int d\vec{r_{1}} \int d\vec{r_{2}} \left(\rho(\vec{r_{1}}) - \rho_{l}\right) \left(\rho(\vec{r_{2}}) - \rho_{l}\right) {\bar{c}}^{(b)}\left(\vec{r_{1}},\vec{r_{2}}\right) \label{2.13} 
\end{align}

 The minimisation is done with an assumed form of \rhor. The ideal part is calculated using a form for \rhor which is a superposition of normalised Gaussians centred around the lattice site,

\begin{align}
\rho(\vec{r}) = \left(\frac{\alpha}{\pi}\right)^{3/2} \sum_{n} \exp \left[-\alpha\left(\vec{r}-\vec{R_{i}}\right)^{2}\right],    \label{2.14}
\end{align}

 where $ \alpha $ is the localization parameter. For the interaction part it is convenient to use Eq(\ref{2.10}). The order parameter $ \rho_{G} = \rho \mu_{G} $ that appears in Eq(\ref{2.10}) is related to parameter $ \alpha $;

\begin{align*}
\mu_{G} = \exp\left[-\frac{G^{2}}{4 \alpha} \right]
\end{align*}

\section{Calculation of $ c^{(0)}(r) $ and $ c^{(b)} \left(\vec{r_{1}},\vec{r_{2}}\right) $}

\subsection{Calculation of $ c^{(0)}(r) $, $ h^{(0)}(r) $ and their derivative with respect to $ \rho $}

 The values of pair correlation functions $ h^{(0)} $ and $ c^{(0)} $ are found from simultaneous solution of the OZ equation,

\begin{align}
h^{(0)}(r) = c^{(0)}(r) + \rho \int d\vec{r'} c^{(0)}(r') h^{(0)}\left(\mid \vec{r'}-\vec{r}\mid\right),   \label{3.1}
\end{align}

 and a closure relation that relates pair correlation functions to pair potential. We use the HMSA (hybridized-mean-spherical approximation) closure of Zerah and Hansen(ZH) \cite{HMSA} which interpolates between the hyper-netted chain (HNC) and soft-core mean spherical approximation (SMSA) relation via a continuous mixing function. The ZH relation is written as

\begin{align}
1+h^{(0)}(r) = \exp\left[-\beta u_{0}(r)\right] ln\left[1+\frac{\exp\left(f(r)\left(\chi^{(0)}(r)-\beta u_{p}(r)\right)\right)-1}{f(r)} \right],     \label{3.2}
\end{align}

 where $ \chi^{(0)}(r) = h^{(0)}(r) - c^{(0)}(r) $, $ f(r) $ is the mixing parameter and $ u_{0}(r) $ and $ u_{p}(r) $ are suitably chosen short-range part and long ranged part of pair potential $ u(r) $. The function $ f(r)= 1 - \exp(-\psi(r)) $ includes an adjustable parameter $ \psi $ which value is chosen to satisfy thermodynamic self consistency between the virial and compressibility routes of the equation of state. This requirement gave us values of $ f(r) $ which are in agreement with those reported in ref. \cite{HMSA} for both systems.

 We used the following two schemes for division of $ u(r) $ of Eq(\ref{1.1}) into $ u_{0}(r) $ and $ u_{p}(r) $. In the WCA scheme (WCAS) $ u_{0}(r) $ is the RLJ potential of Eq(\ref{1.2}) and

\begin{align}
 u_{p}(r)=\begin{cases}
     -\epsilon   & \qquad \qquad r < r_{m} (=2^{1/6}\sigma) \\
     u(r)        & \qquad \qquad r > r_{m},  
     \end{cases}   
\label{3.3}
\end{align}

 In the other scheme referred to as  optimized division scheme (ODS) \cite{ODS} $ u_{p}(r) $ is written as 

\begin{align}
 u_{p}(r)=\begin{cases}
     -p\epsilon   & \qquad r \leq r_{1} \\
     a_{1}+a_{2}r+a_{3}r^{2}+a_{4}r^{3}   & \qquad r_{1} < r \leq r_{2} \\
     u(r)        & \qquad r > r_{2}  
     \end{cases}
\label{3.4}
\end{align}

 and $ u_{0}(r) = u(r) - u_{p}(r) $. Note that for $ p=1 $ and $ r_{1} = r_{2} $ the ODS reduces to the WCAS. The values of $ a_{i} $ parameters are

\begin{align}
\begin{split}
a_{1} &= \dfrac{r_{1}^{3}u(r_{2}) - r_{1}^{3}r_{2}u'(r_{2})-3r_{1}^{2}r_{2}u(r_{2})+r_{1}^{2}r_{2}^{2}u'(r_{2})+3p\epsilon r_{1}r_{2}^{2}-p\epsilon r_{2}^{3}}{(r_{1}-r_{2})^{3}}, \\
a_{2} &=-\dfrac{r_{1}\left(-r_{1}^{2}u'(r_{2})-r_{1}r_{2}u'(r_{2})+6p\epsilon r_{2} -6 r_{1}u(r_{2})+2r_{2}^{2}u'(r_{2})\right)}{(r_{1}-r_{2})^{3}}, \\
a_{3} &=\dfrac{-2r_{1}^{2}u'(r_{2})+3p\epsilon r_{1}-3r_{1}u(r_{2})+r_{1}r_{2}u'(r_{2})+3p\epsilon r_{2}-3r_{2}u(r_{2})+r_{2}^{2}u'(r_{2})}{(r_{1}-r_{2})^{3}}, \\
a_{4} &=-\dfrac{-r_{1}u'(r_{2})-2u(r_{2})+r_{2}u'(r_{2})+2p\epsilon}{(r_{1}-r_{2})^{3}}
\end{split}
\label{3.5}
\end{align}

with $ p=2, r_{1}=0.88\sigma $, $ r_{2} = 1.6\sigma $ and $ u'(r) = \frac{\partial u(r)}{\partial r} $.

For RLJ potential $ u_{p}(r) $ is zero and the ZH closure reduces to the of Roger and Young closure \cite{rogers-PRA-1984}.

The OZ and closure relations for $ \frac{\partial h^{(0)}(r)}{\partial \rho} $ and $ \frac{\partial c^{(0)}(r)}{\partial \rho} $ are found by differentiating Eqs(\ref{3.1}) and (\ref{3.2}) with respect to $ \rho $. Thus

\begin{align}
\dfrac{\partial \chi^{(0)}(r)}{\partial \rho} &= \int d\vec{r'} c^{(0)}(r') h^{(0)}\left(\mid \vec{r'}-\vec{r}\mid\right) \nonumber \\
& \quad + \rho \int d\vec{r'} \dfrac{\partial c^{(0)}(r')}{\partial \rho} h^{(0)}\left(\mid \vec{r'}-\vec{r}\mid\right)  \nonumber \\
& \quad + \rho \int d\vec{r'} c^{(0)}(r') \dfrac{\partial h^{(0)}\left(\mid \vec{r'}-\vec{r}\mid\right)}{\partial \rho}
\label{3.6}
\end{align}

and

\begin{align}
\dfrac{\partial h^{(0)}(r)}{\partial \rho} = e^{-\beta u_{0}(r)}\  \frac{\partial \chi^{(0)}(r)}{\partial \rho}\ e^{f(r)\left(\chi^{(0)}(r)-\beta u_{p}(r)\right)}
\label{3.7}
\end{align}

The closed set of coupled equations (\ref{3.1}),(\ref{3.2})and  (\ref{3.6})-(\ref{3.7})have been solved for four unknowns $ h^{(0)}(r) $, $ c^{(0)}(r) $, $ \frac{\partial h^{(0)}(r)}{\partial \rho} $ and $ \frac{\partial c^{(0)}(r)}{\partial \rho} $  for potentials of Eqs(\ref{1.1}) and (\ref{1.2}).

\begin{figure}[ht]
\includegraphics[height=3.0in, width=4.0in,clip]{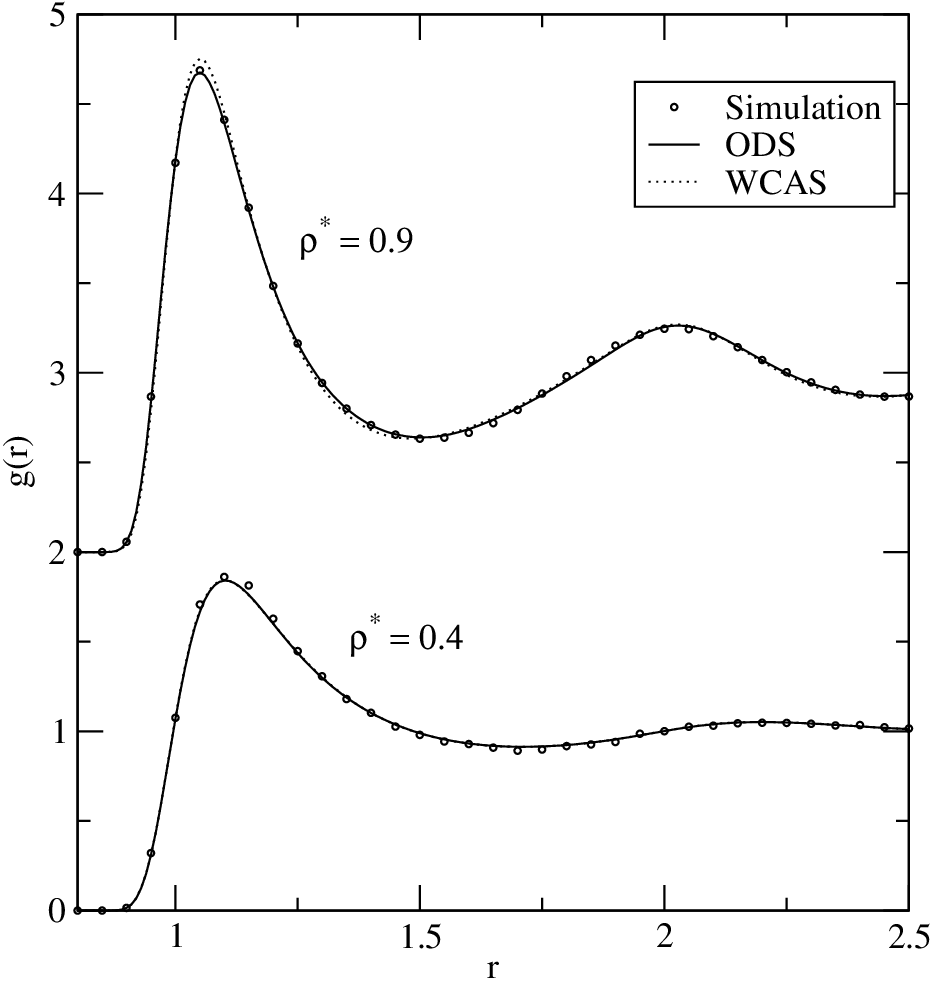}
\caption{Comparison of $ g^{(0)}(r) $ ($ r $ is measured in unit of $ \sigma $) found from WCAS and ODS of division of LJ
potential with the simulation results\cite{llano-JCP-1992} for $ \rho^{*} = 0.4\ \text{and}\ 0.9 $ at $ T^{*} = 1.50 $.}
\label{fig-1}
\end{figure}

\begin{figure}[ht]
\includegraphics[height=3.0in, width=4.0in,clip]{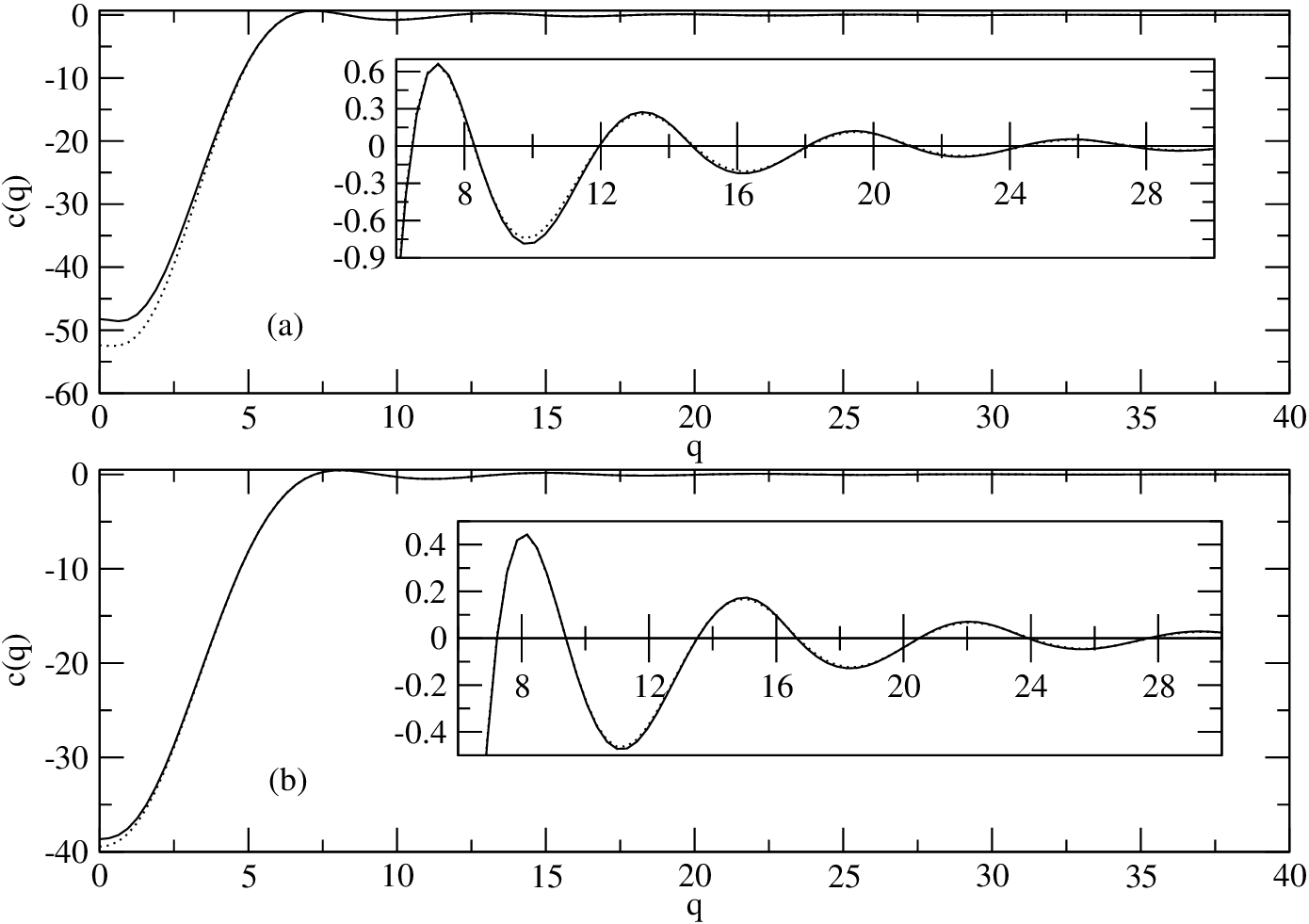}
\caption{Comparison between $ \hat{c}^{(0)}(q) $ found from WCAS and ODS of division of LJ
potential for (a) $ \rho^{*} = 1.05, T^{*} = 1.50\ \text{and\ (b)}\ \rho^{*} = 1.50, T^{*} = 10.0 $. Full and dotted lines correspond to WCAS and ODS respectively.}
\label{fig-2}
\end{figure}

\begin{figure}[ht]
\includegraphics[height=3.0in, width=4.0in,clip]{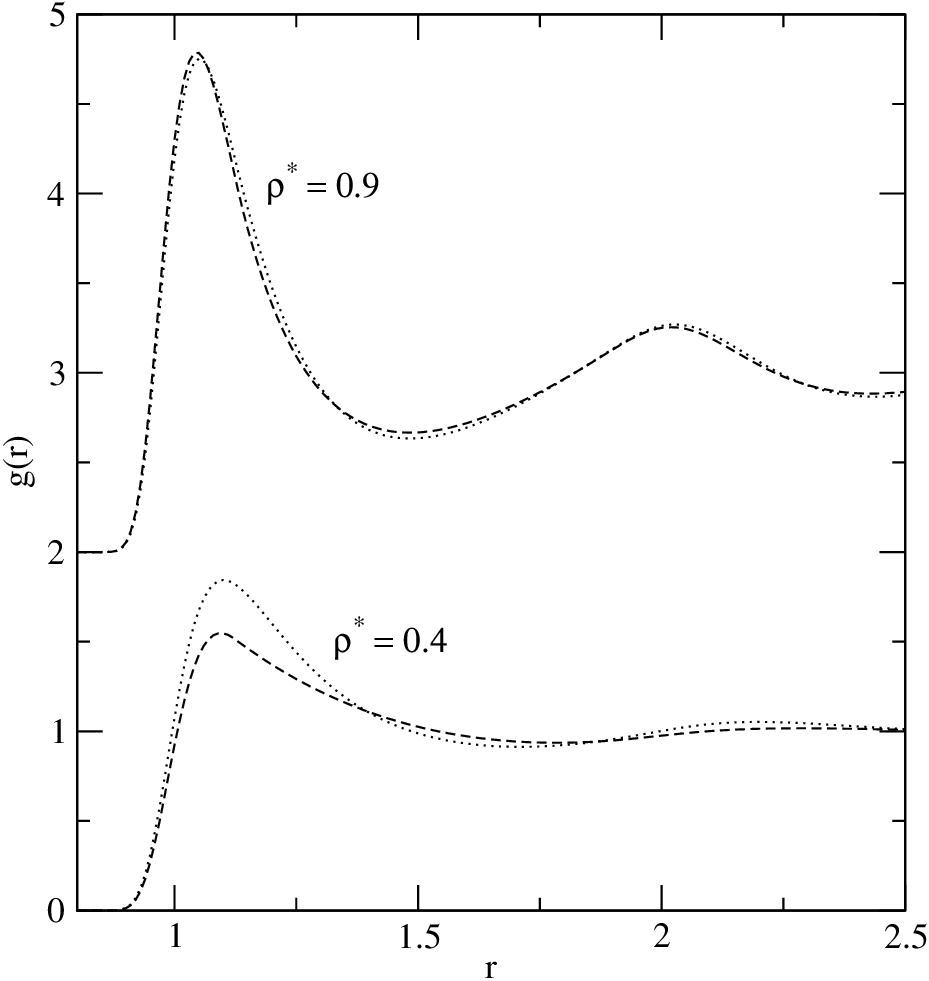}
\caption{Comparison of $ g^{(0)}(r) $ found from LJ potential(WCAS) and RLJ potential for $ \rho^{*} = 0.4\ \text{and}\ 0.9 $ at $ T^{*} = 1.50 $. The dashed and dotted lines are for the RLJ and LJ potentials respectively}
\label{fig-3}
\end{figure}

In Fig.\ref{fig-1} we compare $ g^{(0)}(r) = 1 + h^{(0)}(r) $ found from WCAS and ODS of division of LJ potential with simulation results\cite{llano-JCP-1992} for $ \rho^{*} \ (=\rho\sigma^{3}) = 0.4 $ and $ 0.9 $ at $ T^{*}\ (=k_{B}T/\epsilon) = 1.5 $. As found  in ref. \cite{ODS} the ODS gives better agreement particularly at the first maximum with simulation results than the WCAS. In Fig.\ref{fig-2} we compare $ \hat{c}^{(0)}(q) $ (the Fourier transform of $ c^{(0)}(r) $ ) found from these two schemes for $ \rho^{*}=1.05 $ at $ T^{*}=1.50 $ and $ \rho^{*}=1.50 $ at $ T^{*}=10.0 $ which are close to freezing point. On the scale of the figure the two schemes give almost same values of $ \hat{c}^{(0)}(q) $ except at small value of $ q $. In Table-\ref{CG-DATA} we compare values of $ \hat{c}^{(0)}(G) $ where $ G = \frac{2\pi}{a}\sqrt{l^{2}+m^{2}+n^{2}} $ (l, m, n being integers) are RLV of a fcc lattice and $ a=\left(4/\rho^{*}\right)^{1/3} $ for $ \rho^{*}=1.074 $ at $ T^{*}=1.50 $ and $ \rho^{*}=1.556 $ at $ T^{*}=10.0 $ found from the two schemes. Though the difference in the values of $ \hat{c}(G) $ is small, it has noticeable effect on the freezing parameters as shown below in Figs \ref{fig-9} and \ref{fig-10} and Table-\ref{LJ-Comp}. In Fig.\ref{fig-3} we compare values of $ g^{(0)}(r) $ of LJ potential with that of RLJ potential at $ \rho^{*}=0.4 $ and $ 0.9 $ for $ T^{*}=1.50 $. The two values are in good agreement at high density but differ at lower density; this is because of the contribution of attractive interaction which decreases with increasing density.

\subsection{Calculation of $ c^{b}(\vec{r_{1}},\vec{r_{2}}) $}

One can use the relation 

\begin{align}
\dfrac{\delta^{n-2} c(\vec{r_{1}},\vec{r_{2}})}{\delta \rho(\vec{r_{3}}) \cdots \cdots \delta \rho(\vec{r_{n}})} = c_{n}(\vec{r_{1}},\vec{r_{2}},\cdots \cdots,\vec{r_{n}}),
\label{3.8}
\end{align}

where $ c_{n} $ is the n-body direct correlation function (DPF) and the functional Taylor expansion to write the following series for $ c^{b}(\vec{r_{1}},\vec{r_{2}}) $.

\begin{align}
c^{(b)}(\vec{r_{1}},\vec{r_{2}}) &=\int d\vec{r_{3}} c_{3}^{(0)} (\vec{r_{1}},\vec{r_{2}},\vec{r_{3}};\rho_{s}) (\rho(\vec{r_{3}})-\rho_{s}) \nonumber \\
&\quad+\frac{1}{2} \int d\vec{r_{4}} c_{4}^{(0)} (\vec{r_{1}},\vec{r_{2}},\vec{r_{3}},\vec{r_{4}};\rho_{s}) (\rho(\vec{r_{3}})-\rho_{s})(\rho(\vec{r_{4}})-\rho_{s}) \nonumber \\
&\quad+ \cdots
\label{3.9}
\end{align}

In Eq(\ref{3.9}) $ c_{m}^{(0)} $ is the m-body DCF of a homogeneous system of density \rhos and  $ \rho(\vec{r})-\rho_{s} = \sum_{G} \rho_{G} \exp(i\vec{G} \cdot \vec{r})$. The values of $ c_{m}^{(0)} $ can be found from exact relations

\begin{align}
\dfrac{\partial^{n}c^{(0)}(r;\rho)}{\partial \rho^{n}} = \int d\vec{r_{3}} \cdots \int d\vec{r_{n}} c_{n+2}^{(0)}(\vec{r_{1}},\vec{r_{2}},\cdots \cdots,\vec{r_{n+2}}).
\label{3.10}
\end{align}

The values of $ \frac{\partial^{n}c^{(0)}(r;\rho)}{\partial \rho^{n}} $ and the factorization \textit{ansatz} can be used to find values of $ c_{n+2}^{(0)} $ from Eq(\ref{3.10}). The factorization \textit{ansatz} which was first used by Barrat et al \cite{barrat-MOLPHYS-1988} to calculate $ c_{3}^{(0)} $ has recently been extended by Bharadwaj et al \cite{bharadwajas-PRE-2013} to calculate  $ c_{4}^{(0)} $.

In the case of inverse power potential it was found that at the melting point  $ c^{(b)} $ is accurately approximated by the first term of series (\ref{3.9}) even for very soft repulsions\cite{bharadwajas-PRE-2013}; the contribution made by $ c^{(b)} $ to free energy increases with the range of the potential. Since, as shown below, the contribution made by the attractive part of the LJ potential at the transition point is small and contribute opposite to that of the repulsive part, we expect the conclusion drawn in case of the inverse power potentials holds in the present systems as well. In view of this, we consider the first term of series (\ref{3.9}) and examine its effect on the freezing parameters. Following Barrat et al \cite{barrat-MOLPHYS-1988} we write

\begin{align}
c_{3}^{(0)}(\vec{r_{1}},\vec{r_{2}},\vec{r_{3}}) = t(r_{12}) t(r_{13}) t(r_{23})
\label{3.11}
\end{align}

and determine the function $ t(r) $ from the relation 
\begin{align}
\dfrac{\partial c^{(0)}(r;\rho)}{\partial \rho} =  t(r) \int d\vec{r'} \ t(r') \ t\left(\mid \vec{r'} - \vec{r} \mid\right)
\label{3.12}
\end{align}
using an iterative procedure. From known values of $ t(r), c_{3}^{(0)} $ is found from Eq(\ref{3.11}). It was shown in ref \cite{barrat-MOLPHYS-1988} that the value of $ c_{3}^{(0)} $ calculated in this way for the inverse power potential agrees with simulation results. It may also be shown that $ \hat{c}^{(0)}(\vec{q_{1}}, \vec{q_{2}}, \vec{q_{3}}) $ agrees with exact three-body DCF at least up to the second order in the wave numbers.

Using Eq(\ref{3.11}) in the first term of the series (\ref{3.9}) and substituting the value of $ \rho(\vec{r_{3}}) - \rho_{s} $ we find

\begin{align}
c^{(b)}(\vec{r_{1}},\vec{r_{2}}) = \sum_{G} \ e^{i\vec{G}\cdot \vec{r_{c}}}\ t(r)\ e^{-\frac{1}{2}i \vec{G}\cdot\vec{r}} \int d\vec{r'}\ t(r')\ t(\mid\vec{r'}-\vec{r}\mid)\ e^{i\vec{G}\cdot\vec{r'}}
\label{3.13}
\end{align}
where $ \vec{r} = \vec{r_{2}}-\vec{r_{1}} $, $ \vec{r_{c}} = \frac{1}{2}(\vec{r_{1}}+\vec{r_{2}}) $ and $ \vec{r'} = \vec{r_{3}}-\vec{r_{1}} $

This is solved to give \cite{singhsl-EPL-2009,bharadwajas-PRE-2013,singhsl-PRE-2011}
\begin{align}
c^{(b)}(\vec{r_{1}},\vec{r_{2}}) = \sum_{G} \ e^{i\vec{G}\cdot \vec{r_{c}}}\ \sum_{l m} c_{l}^{(G)}(r) Y_{lm}^{*}(\hat{G}) Y_{lm}(\hat{r})
\label{3.14}
\end{align}

where
\begin{align}
c_{l}^{(G)}(r) = \rho_{G}\sum_{l_{1}}\sum_{l_{2}}\Lambda(l,l_{1},l_{2}) j_{l_{2}}\left(\frac{1}{2}Gr\right)\ B_{l_{1}}(r,G)
\label{3.15}
\end{align}

Here $ j_{l}(x) $ is the spherical Bessel function, $ Y_{lm}(\hat{x}) $ the spherical harmonics,

\begin{align*}
\Lambda(l,l_{1},l_{2}) = (i)^{l_{1}+l_{2}}(-1)^{l_{2}} \left[\frac{(2l_{1}+1)(2l_{2}+1)}{2l+1}\right]\left[C_{g}\left(l_{1},l_{2},l;0,0,0\right)\right]^{2}
\end{align*}
and
\begin{align*}
B_{l_{1}}(r,G) = 8\ t(r) \int dk k^{2}\ t(k) j_{l_{1}}(kr)\ \int dr' r^{2}\ t(r') j_{l_{1}}(kr') j_{l_{1}}(Gr')
\end{align*}
where $ C_{g} $ is the Clebsh-Gardon coefficient. The crystal symmetry dictates that $ l $ and $ l_{1} + l_{2} $ are even and for cubic crystal $ m = 0, \pm4 $.

\begin{figure}[ht]
\includegraphics[height=3.0in, width=5.0in,clip]{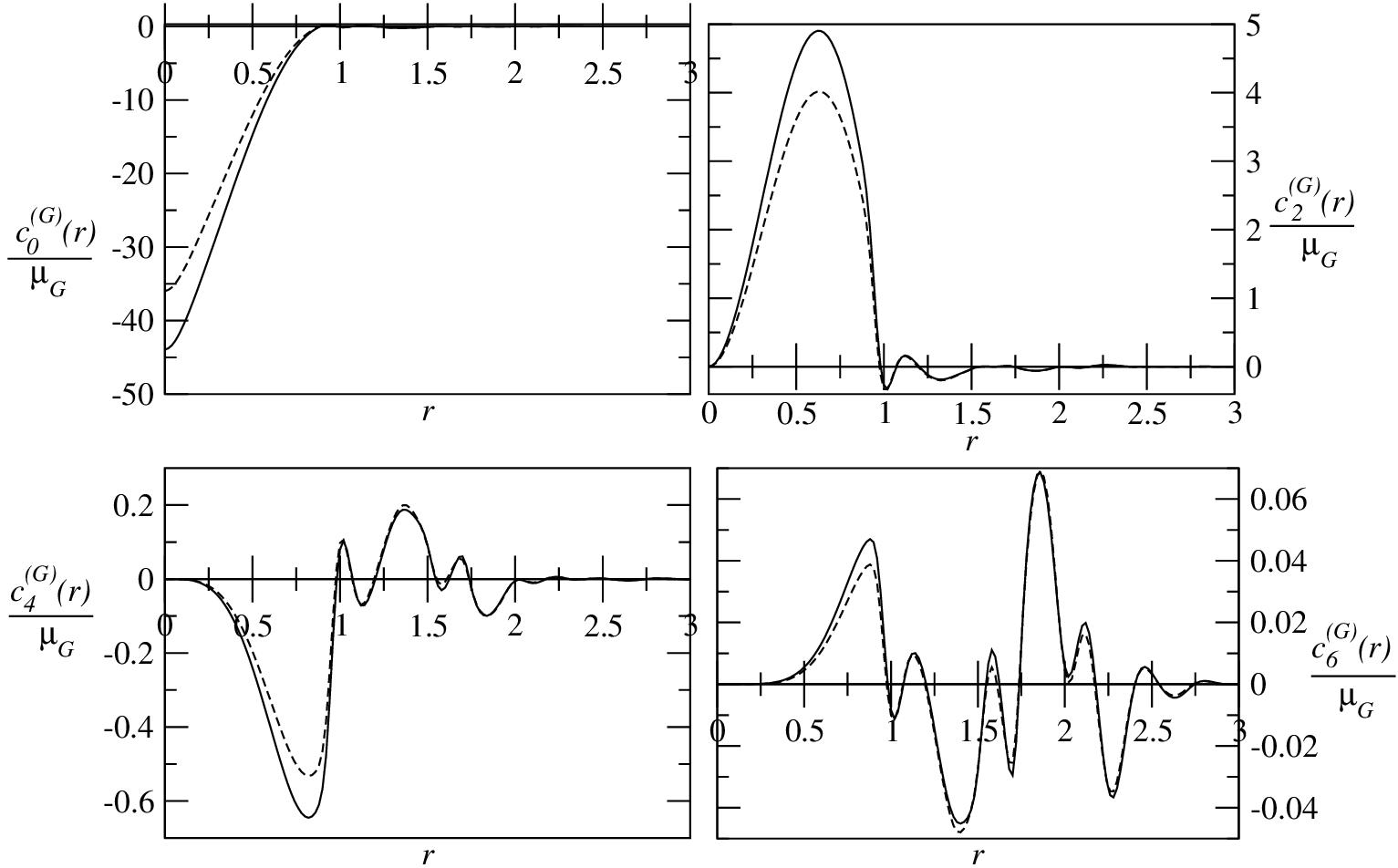}
\caption{Comparison of values of $ \frac{c_{l}^{(G)}(r)}{\mu_{G}} $ as a function of $ r $ (measured in unit of $ \sigma $) for $ l=0,2,4\ \text{and}\ 6 $ for RLV's of first set of a fcc lattice at $ \rho_{s}^{*}= 1.10$ and $ T^{*}=1.50 $; the full and dashed lines correspond to LJ (found using WCAS) and RLJ potentials, respectively.}
\label{fig-4}
\end{figure}

\begin{figure}[ht]
\includegraphics[height=3.0in, width=5.0in,clip]{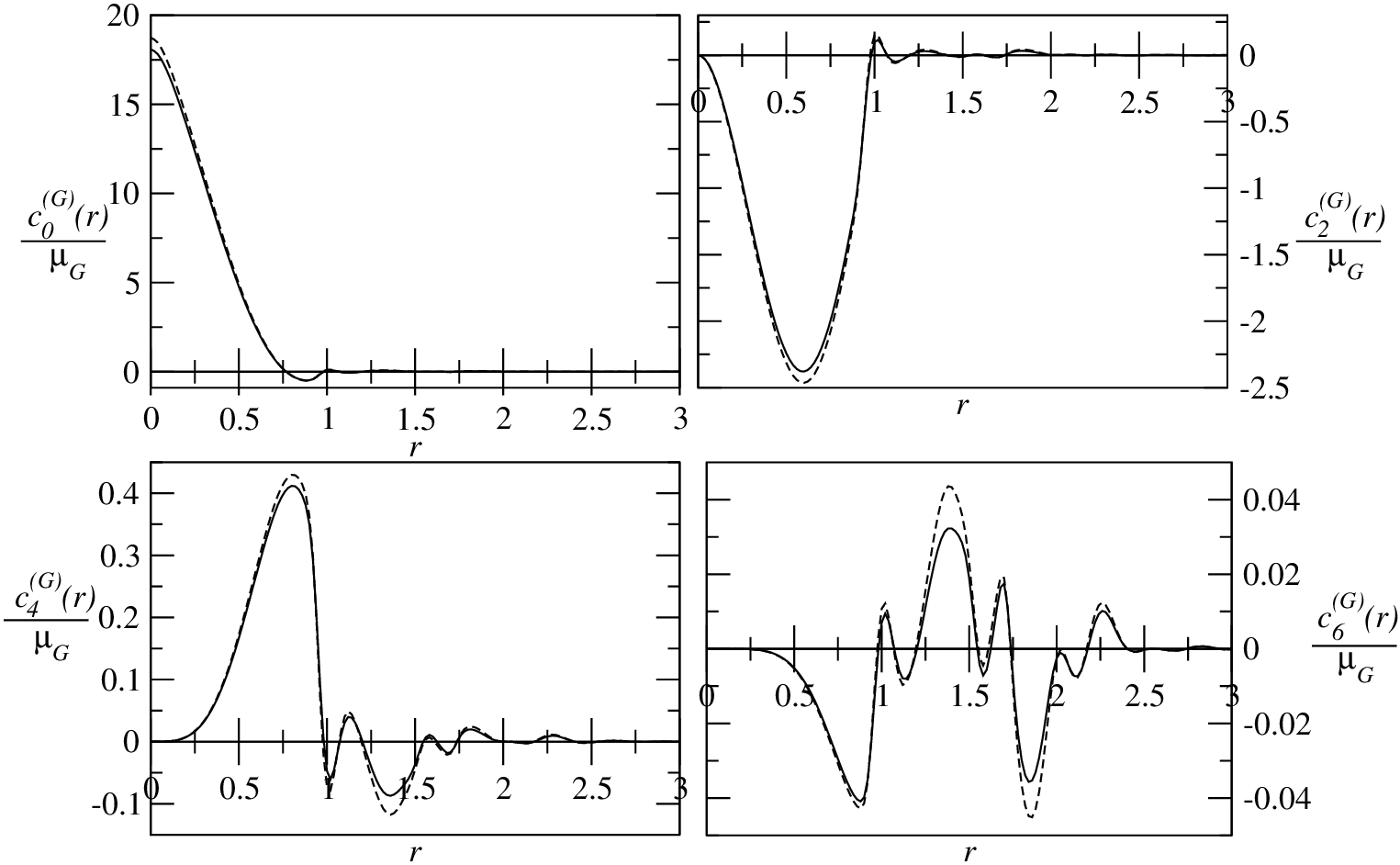}
\caption{Same as for Fig.\ref{fig-4} but for RLV's of second set of a fcc lattice.}
\label{fig-5}
\end{figure}

\begin{figure}[ht]
\includegraphics[height=3.0in, width=5.0in,clip]{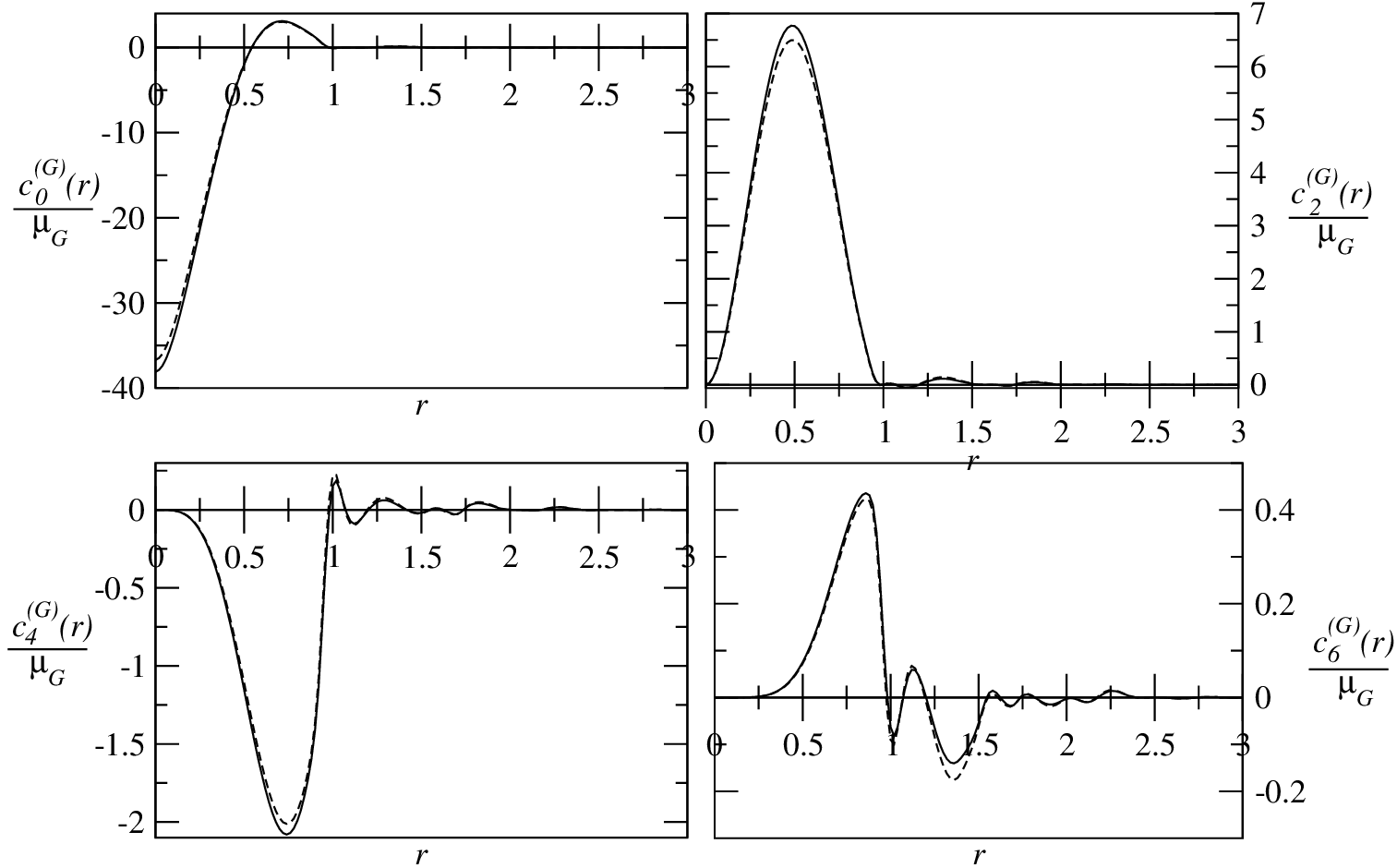}
\caption{Same as for Fig.\ref{fig-4} but for RLV's of third set of a fcc lattice.}
\label{fig-6}
\end{figure}

The values of $ c_{l}^{(G)}(r) $ depend on order parameters \muG and on magnitude of $ \vec{G} $. In Figs \ref{fig-4}-\ref{fig-6}\ we plot and compare $ \frac{c_{l}^{(G)}(r)}{\mu_{G}} $ for $ l=0,2,4\ \text{and}\ 6 $ for RLV's of first three sets, respectively, of a fcc lattice at $ \rho_{s}^{*}= 1.10$ and $ T^{*}=1.50 $; the full and dashed lines correspond to LJ (found using WCAS) and RLJ potentials respectively. For a different set of RLV's $ \frac{c_{l}^{(G)}(r)}{\mu_{G}} $ varies with $ r $ in different ways. The values in all cases become negligible for $ r\ (r\ \text{is measured in unit of}\ \sigma) > 2.5 $. For any given $ G $, \ $ \frac{c_{l}^{(G)}(r)}{\mu_{G}} $ decreases rapidly with $ l $; major contribution comes from $ l=0\ \text{and}\ 2$. For $ l=6 $ the value is about three order of magnitude smaller than that of $ l=0 $. It is also seen that at any given point $ r $, values of $ \frac{c_{l}^{(G)}(r)}{\mu_{G}} $ are positive for some set of $ G $  while for other values are negative, leading to mutual cancellation in a quantity where summation over $ G $ is involved. The difference between the values of $ \frac{c_{l}^{(G)}(r)}{\mu_{G}} $ of LJ and RLJ is maximum for the first set of $ \vec{G} $ vectors and becomes almost negligible for other sets, showing the limited effect that the attractive interaction has on crystal structure.

\subsection{Calculation of $ \bar{c}^{(0)}(r) $ and $ \bar{c}^{(b)}(\vec{r_{1}},\vec{r_{2}}) $}

As shown in ref \cite{jaiswala-PRE-2013, bharadwajas-PRE-2013},  $ \bar{c}^{(0)}(r;\rho) $ can be approximated as 

\begin{align*}
\bar{c}^{(0)}(r;\rho) = {c}^{(0)}(r;\rho_{l}) + \frac{1}{3} \rho_{l} \Delta \rho^{*} \frac{\partial {c}^{(0)}(r;\rho_{l})}{\partial \rho_{l}}
\end{align*}
where the contribution arising from the second term to the free energy is found to be negligibly small and one can replace $ \bar{c}^{(0)}(r;\rho) $ by $ {c}^{(0)}(r;\rho_{l}) $.

For evaluation of $ \bar{c}^{(b)} $, we note that it is linear in order parameter and the integration over $ \xi $ variable in Eq(\ref{2.9}) can be performed analytically  leading to

\begin{align*}
\bar{c}^{(b)}(\vec{r_{1}},\vec{r_{2}}) = \sum_{G} \ e^{i\vec{G}\cdot \vec{r_{c}}}\ \sum_{l m} \bar{c}_{l}^{(G)}(r) Y_{lm}^{*}(\hat{G}) Y_{lm}(\hat{r})
\end{align*}

where
\begin{align*}
\bar{c}_{l}^{(G)}(r) = \rho_{G}\sum_{l_{1}}\sum_{l_{2}}\Lambda(l,l_{1},l_{2}) j_{l_{2}}\left(\frac{1}{2}Gr\right)\ \bar{B}_{l_{1}}(r,G)
\end{align*}
with
\begin{align*}
\bar{B}_{l_{1}}(r,G) = 2\ \int_{0}^{1} d\lambda\ (1-\lambda)\  B_{l_{1}}(r,G;\lambda\rho)
\end{align*}

The quantity $ B_{l_{1}}(r,G) $ is defined by Eq(\ref{3.15}). The integration over $ \lambda $ has been performed numerically by varying it from $ 0 $ to $ 1 $ on a fine grid and evaluating $ B_{l_{1}} $ on these densities. Since this function vary smoothly with density and its value has been evaluated at closely spaced values of density, the result for $ \bar{c}^{(b)}(\vec{r_{1}},\vec{r_{2}}) $ is expected to be accurate.

\begin{figure}[ht]
\includegraphics[height=3.0in, width=6.0in,clip]{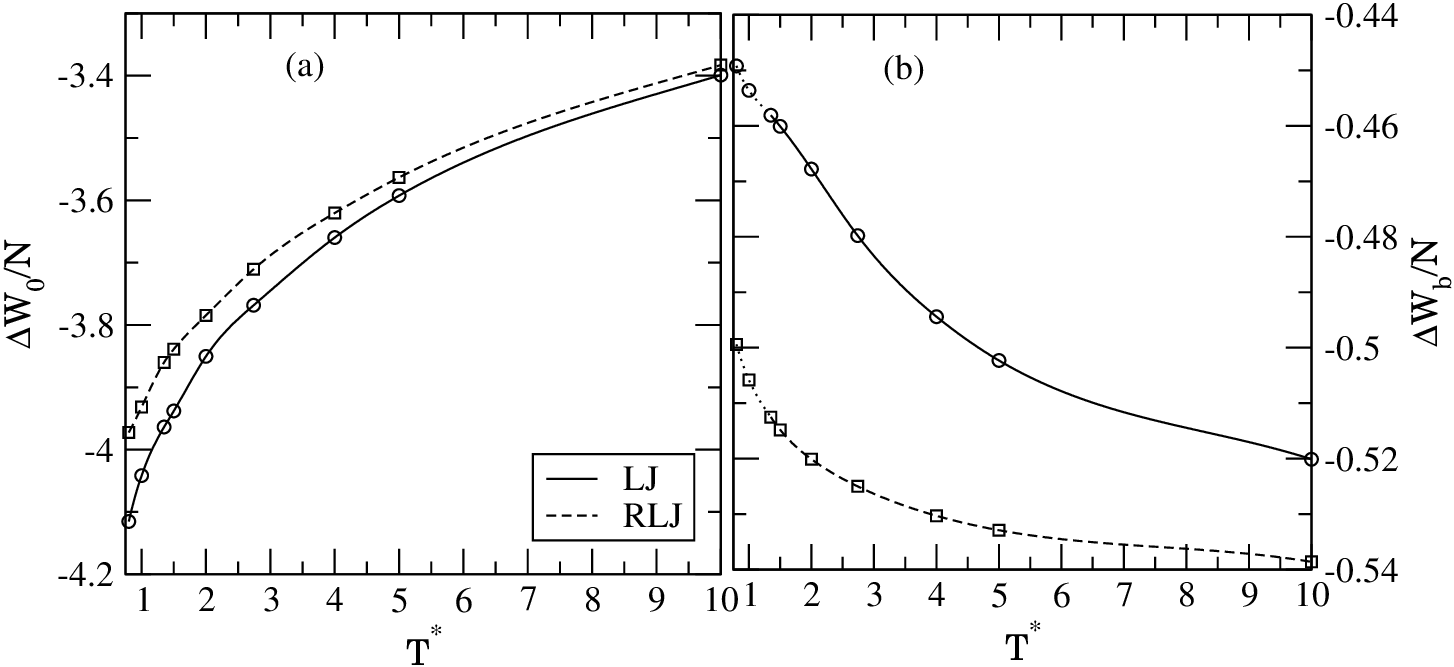}
\caption{Comparison of values of $ \frac{\Delta W_{0}}{N}  $ and $ \frac{\Delta W_{b}}{N} $ at the freezing points for the LJ (circles) and RLJ (squares) potentials. The dotted part of lines in (b) represent extrapolated values of $ \frac{\Delta W_{b}}{N} $.}
\label{fig-11}
\end{figure}

As noted in ref \cite{kuijper-JCP-1990}, the HMSA closure dose not give self-consistent solutions for these potentials at low densities and low temperatures ($ \rho^{*} \leq 0.5 $ and $ T^{*} \leq 1.0 $) we could not calculate accurately the value of $ \bar{c}^{(b)}(\vec{r_{1}},\vec{r_{2}}) $ below the critical temperature ($ T_{c}^{*} \simeq 1.35 $) for the LJ potential and for $ T^{*} \leq 1 $ for the RLJ potential. Below these temperatures we have therefore used extrapolated values of free energy contribution due to symmetry broken part of DPCF (see Fig \ref{fig-11}) to locate the freezing transition.

\section{Liquid-Solid Transition}

From Eqs (\ref{2.12})-(\ref{2.13}) and expressions for $ \bar{c}^{(0)}(r) $ and $ \bar{c}^{(b)}(\vec{r_{1}},\vec{r_{2}}) $ given above one finds \cite{singhsl-EPL-2009,bharadwajas-PRE-2013,singhsl-PRE-2011}

\begin{align}
\frac{\Delta W}{N} = \frac{\Delta W_{id}}{N} + \frac{\Delta W_{0}}{N} + \frac{\Delta W_{b}}{N}
\label{4.1}
\end{align}
where
\begin{align}
\frac{\Delta W_{id}}{N} &= 1 - (1 + \Delta \rho^{*})\left[\frac{5}{2} + ln\rho_{l}^{*} - \frac{3}{2} ln \left(\frac{\alpha}{\pi}\right)\right], \label{4.2} \\ 
\frac{\Delta W_{0}}{N} &= -\frac{1}{2}\ \rho_{l}^{*} {\Delta\rho^{*}}^{2} \bar{c}^{(0)}(0) - \frac{1}{2}\ \rho_{l}^{*} (1 + \Delta \rho^{*})^{2}\ \sum_{G\neq0}\mid\mu_{G}\mid^{2}\ \bar{c}^{(0)}(G), \label{4.3}  \\
 \frac{\Delta W_{b}}{N} &= -\frac{1}{2}\ \rho_{l}^{*} (1 + \Delta \rho^{*})^{2}\ {\sum_{G}}^{'} {\sum_{G_{1}}}^{'} \mu_{G_{1}} \mu_{-G-G_{1}}\ \bar{c}^{(G)}\left(\vec{G_{1}}+\frac{1}{2}\vec{G}\right) \label{4.4}
\end{align}

Here $ \Delta W_{id} $, $ \Delta W_{0} $ and $ \Delta W_{b} $ are respectively, the ideal, the symmetry-conserving and the symmetry broken contributions to $ \Delta W $. The prime on summation in Eq(\ref{4.4}) indicates the condition $ \vec{G} \neq 0 $, $ \vec{G_{1}} \neq 0 $ and $ \vec{G_{1}} \neq \vec{G} $ and

\begin{align}
\hat{c}^{(0)}(G) = 4\pi \int dr \ r^{2}\ c^{(0)}(r)\ j_{0}(Gr),
\label{4.5}
\end{align}

\begin{align}
\hat{\bar{c}}^{(G)}\left(\vec{A}\right) = 4\pi \sum_{l m} i^{l}\ \int dr \ r^{2}\ c_{l}^{(G)}(r)\ j_{l}\left(Ar\right) Y_{lm}(\hat{A}),
\label{4.6}
\end{align}
where $ \vec{A} = \left(\vec{G_{1}}+\frac{1}{2}\vec{G}\right) $.

These equations are used to locate the fluid-fcc crystal transition. The reason for selecting the fcc structure are following; (i) these systems are known to freeze into fcc crystal, (ii) simulation data are mostly for fluid-fcc crystal transition \cite{agrawal-MOLPHYS-1988, ahmed-JCP-2009, sousa-JCP-2012, hansen-PRev-1969, *hansen-PRA-1970,ahmed-PRE-2009-WCA,kuijper-JCP-1990-WCA} and (iii) the difference between freezing density of fcc lattice and hexagonal closed packed (hcp) lattice is very small (the hcp density is slightly higher). The $ \frac{\Delta W}{N} $ is minimized with respect to two parameters $ \rho_{s}^{*} $ and $ \alpha $. For a given $ \rho_{s}^{*} $ and $ \Delta \rho^{*} $, $ \frac{\Delta W}{N} $ is minimised with respect to $ \alpha $; next $ \Delta \rho^{*} $ is varied till the lowest value of $ \frac{\Delta W}{N} $ at its minimum is found. If this lowest value of $ \frac{\Delta W}{N} $ is not zero then $ \rho_{s}^{*} $ is varied until $ \frac{\Delta W}{N} $ is zero. The lowest value of $ \rho_{s}^{*} $ and corresponding  $ \rho_{l}^{*} = \rho_{s}^{*}/(1+\Delta \rho^{*}) $ for which the condition $ \frac{\Delta W}{N} = 0$ is satisfied are taken as the coexisting solid and fluid densities at the transition. This procedure has been used in finding values of freezing parameters from the present theory (Eqs (\ref{4.1}) - (\ref{4.6})) as well as from the RY-DFT.

In Table-\ref{RLJ-Comp} we compare values of freezing parameters $ \rho_{l}^{*} $, $ \rho_{s}^{*} $, $ \Delta \rho^{*} $, the Lindemann parameter $ L $ and $ P^{*} = P\sigma^{3}/\epsilon $, where $ P $ is the pressure at the freezing point, found from our theory with those found from the RY-DFT, MWDA \cite{kuijper-JCP-1990} and simulations \cite{ahmed-PRE-2009-WCA, kuijper-JCP-1990-WCA} for the RLJ potential. The RY-DFT gives values of $ \rho_{l}^{*} $ and $ P^{*} $ which are quite high compared to simulation values, e.g. at $ T^{*}=2 $, $ \rho_{l}^{*} $ is about $ 9\% $ and $ P^{*} $ is about $ 34\% $ higher. The MWDA while gives relatively better agreement at higher temperatures, fails at low temperatures. The values found from our theory, (given in the first row of the table) are in very good agreement with  simulation results for the entire temperature range.

\begin{figure}[ht]
\includegraphics[height=3.0in, width=4.0in,clip]{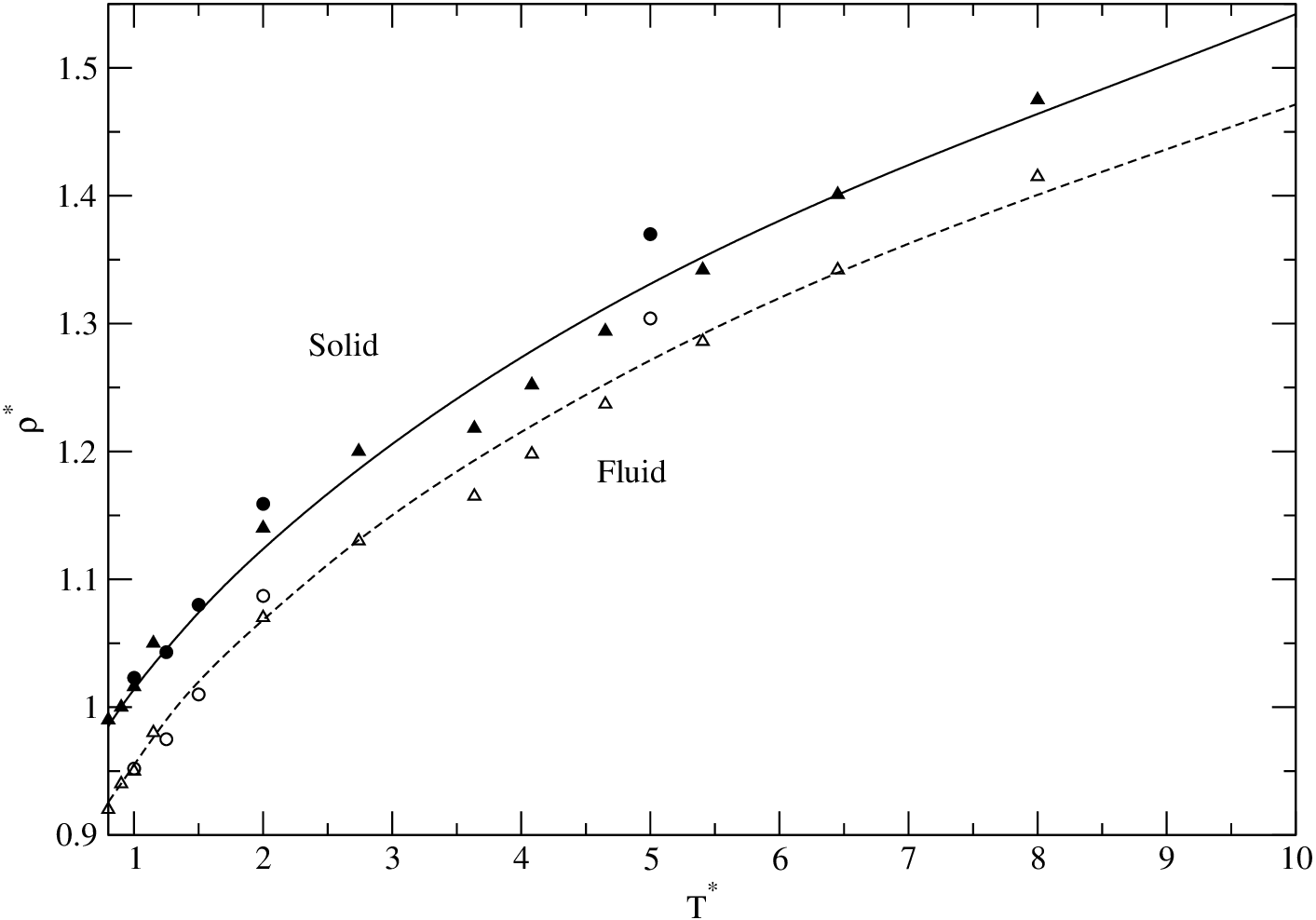}
\caption{The $ T^{*}-\rho^{*} $  phase diagram for the RLJ potential. Full and dashed lines correspond respectively to crystal and fluid at the freezing point, triangles and circles (full for crystal and open for fluid) repent simulation data of refs. \cite{ahmed-PRE-2009-WCA} and \cite{kuijper-JCP-1990-WCA} respectively.} 
\label{fig-7}
\end{figure}

\begin{figure}[ht]
\includegraphics[height=3.0in, width=4.0in,clip]{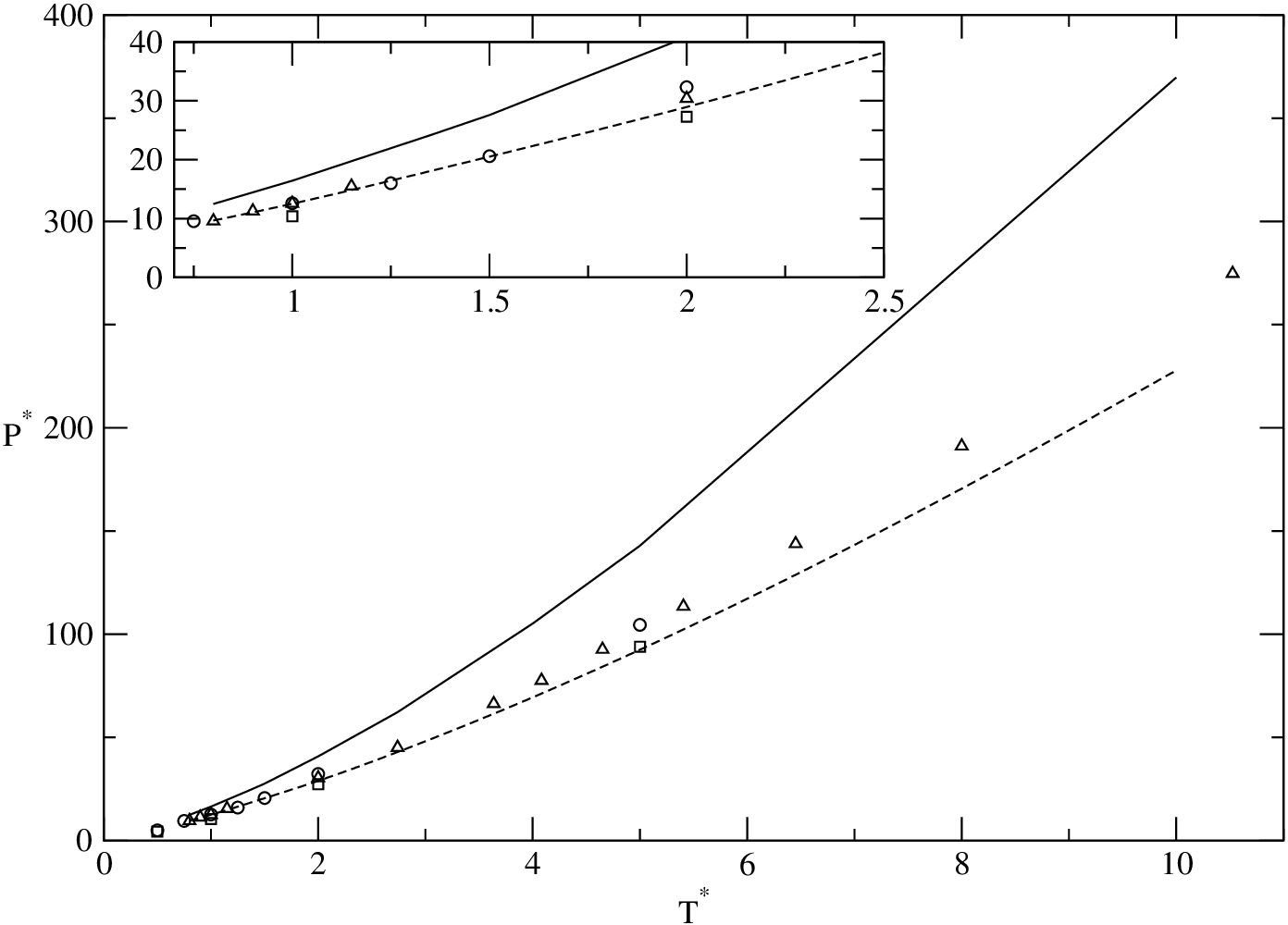}
\caption{Pressure $P^{*}$ \textit{vs} temperature $ T^{*} $  for RLJ potential. Dashed line represents present data and full line represents data found from RY-DFT, squares represent values from MWDA (\cite{kuijper-JCP-1990}) triangles and circles represent simulation data of refs. \cite{ahmed-PRE-2009-WCA} and \cite{kuijper-JCP-1990-WCA} respectively.}
\label{fig-8}
\end{figure}

In Fig.\ref{fig-7} we plot the solid - fluid phase diagram; the lines (full line for fcc crystal and dashed line for fluid) are from the present theory and circles and squares ( open for fluid and full for crystal) are from simulations \cite{ahmed-PRE-2009-WCA, kuijper-JCP-1990-WCA}. We note large spread in simulation values. This may be due to different theoretical methods used in locating the transition and system sizes in the calculation. One may also note the values given in ref \cite{ahmed-PRE-2009-WCA} for low temperatures ($ T^{*} \leq 2.74 $) and high temperatures ($ T^{*} \geq 3.63636 $) do not seem to join smoothly. This may be due to use of two different algorithms in these two temperature regions. In Fig.\ref{fig-8} we plot $ P^{*} $ vs $ T^{*} $, dashed line from present theory, full line from RY-DFT and open circles and triangles from simulations and squares from MWDA.

In Table-\ref{LJ-Comp} we compare the values of freezing parameters for the LJ potential. The values found from ODS and WCAS of division of potential into reference and perturbation are also compared. It may be noted that while the values of $ \rho_{l}^{*} $, $ \rho_{s}^{*} $ and therefore $ P^{*} $ found from ODS are somewhat higher but $ \Delta \rho^{*} $ is lower  than those found from WCAS, This is because of the difference in the values of $ \hat{c}(q) $ shown in Fig.\ref{fig-2} and Table-\ref{CG-DATA}. As in the case of RLJ potential, the values found from RY-DFT for $ \rho_{l}^{*} $, $ \rho_{s}^{*} $ and $ P^{*} $ are quite high compared to simulation values. The MWDA, as shown in ref. \cite{kuijper-JCP-1990} did not yield a (meta-) stable solid phase at $ T^{*} < 5.00 $. However, at $ T^{*} = 10.0 $ the theory gave values which are in good agreement with simulation results. 

The solid-fluid phase diagram is plotted in Fig.\ref{fig-9}. The simulation values given in the table and in the figure are of \citet{agrawal-MOLPHYS-1988}, \citet{ahmed-JCP-2009}, \citet{sousa-JCP-2012} and \citet{hansen-PRev-1969,*hansen-PRA-1970}. The large spread in the simulation values is seen in this case also. While both the ODS and WCAS results are in good agreement with simulation results, the ODS values are in better agreement with simulation values at high temperatures $ T^{*} > 2.0 $ whereas WCAS values are closer to simulation values for $ T^{*} < 2.0 $. The value of Lindemann parameter (a measure of the relative displacement of particle around its lattice position) found by both methods is almost same and varies marginally with temperature; e.g. it varies from $ 0.092 $ at $ T^{*}=0.8 $ to $ 0.107 $ at $ T^{*}=10.0 $. In Fig.\ref{fig-10} $ P^{*} $ and $ T^{*} $ is plotted and compared with simulation and RY-DFT results.

\section{Summery and Conclusions}

The free energy functional proposed by Singh and Singh \cite{singhsl-EPL-2009} for a crystal is used to calculate freezing parameters of simple systems interacting via the LJ and the RLJ   potentials. This free energy functional is exact and involves the symmetry conversing part of the DPCF, $ c^{(0)}(r,\rho) $ and the symmetry broken part,  $ c^{(b)}(\vec{r_{1}} , \vec{r_{2}}) $ as input informations. The values of $ c^{(0)}(r) $ which corresponds to isotropy and homogeneity of the phase are found from the integral equation theory comprising the OZ equation and the ZH closure relation\citep{HMSA}. For  $ c^{(b)}(\vec{r_{1}}, \vec{r_{2}}) $, which is a functional of \rhor and is invariant only under a discrete set of translations and rotations, an expansion in ascending powers of order parameters has been used. This expansion involves higher body direct correlation functions of isotropic systems at average density of the crystal \rhos, which in turn were found from the density derivatives of $ c^{(0)}(r) $ using a method describe in refs.\cite{singhsl-EPL-2009,singhsl-PRE-2011,jaiswala-PRE-2013,bharadwajas-PRE-2013}.

Through the contribution of symmetry broken part of DPCF to the grand thermodynamic potential is small compared to the symmetry conserving part, it plays crucial role in freezing of fluids. In Table-\ref{RLJ-Data} we compare the contribution made by the ideal gas part, $ \frac{\Delta W_{id}}{N} $, the symmetry conserving part, $ \frac{\Delta W_{0}}{N} $, and the symmetry broken part, $ \frac{\Delta W_{b}}{N} $ at the freezing point for both potentials at different temperatures. As $ \frac{\Delta W_{b}}{N} $ is negative it adds to $ \frac{\Delta W_{0}}{N} $ to overcome the positive contribution of $ \frac{\Delta W_{id}}{N} $ in order to make $ \frac{\Delta W}{N} = 0$. We note that the contribution of $ \frac{\Delta W_{b}}{N} $, compared to $ \frac{\Delta W_{0}}{N} $, increases with the temperature; albeit marginally. For example it increases from $ 12.6\% $ at $ T^{*} = 0.8 $ to $ 16.0\% $ at $ T^{*} = 10.0 $ for RLJ potential and from $ 11.0\% $ at $ T^{*} = 0.8 $ to $ 15.3\% $ at $ T^{*} = 10.0 $ for the LJ potential. We also note that at the same temperature the relative contribution of $ \frac{\Delta W_{b}}{N} $ for LJ potential is marginally lower than that for RLJ potential.  In Fig-\ref{fig-11} the values of $ \frac{\Delta W_{0}}{N} $ and $ \frac{\Delta W_{b}}{N} $ at the freezing point for these two potentials as a function of temperature are compared. One may note that while attractive interaction contribution to $ \frac{\Delta W_{0}}{N} $ is to increase its value, it decreases the value of $ \frac{\Delta W_{b}}{N} $. This shows that the contribution of attractive interaction to $ \frac{\Delta W_{b}}{N} $ is small and opposite to that of repulsive potential part of interaction. However, these contributions are small leading to conclusion that freezing is predominately determined by the repulsive part of the interaction.

The difference in the values of freezing parameters for the LJ potential found from ODS and WCAS shows that the value of freezing parameters are sensitive to values of DPCF.

In conclusion, we wish to emphasize that the agreement between theory and simulation values of freezing parameters for potentials studied here and elsewhere\cite{singhsl-EPL-2009, jaiswala-PRE-2013, bharadwajas-PRE-2013, singhsl-PRE-2011} shows that the free energy functional proposed by Singh and Singh\cite{singhsl-EPL-2009} provides an accurate theory for fluid - solid transition for a wide class of potentials. As this free energy functional takes into account the spontaneous symmetry breaking, it can be used to study solid-solid transitions as well as other properties of crystals.
\section*{Acknowledgements}
We are thankful to the Department of Science and Technology (DST), University Grants Commission (UGC) and Indian National Science Academy for financial support.
\bibliographystyle{apsrev4-1}
\bibliography{man}
\newpage
\begin{center}
\begin{scriptsize}
\begin{table*}
\caption{Comparison of values of $\ \hat{c}(\mid\vec{G}\mid ) $ where  $\ \mid\vec{G}\mid = \frac{2\pi}{a}\sqrt{l^{2}+m^{2}+n^{2}} $ ($ l,\ m,\ n $ being integers) are RLV of a fcc lattice and $ a = \left( 4/\rho^{*}\right)^{1/3} $ found from ODS and WCAS of division of the LJ potential at $ T^{*} = 1.50,\ \rho^{*} = 1.05 $ and $ T^{*} = 10.0,\ \rho^{*} = 1.50 $.}
\vspace{0.5cm}
\label{CG-DATA}
\begin{ruledtabular}
\begin{tabular}{>{\scriptsize}c>{\scriptsize}c>{\scriptsize}c>{\scriptsize}c>{\scriptsize}c>{\scriptsize}c>{\scriptsize}c>{\scriptsize}c>{\scriptsize}c}
\small
 { S.N.} & & \mc{3}{c}{$ T^{*} = 1.50,\ \rho^{*} = 1.05 \ \text{and}\ a = 1.55 $} & & \mc{3}{c}{$ T^{*} = 10.0,\ \rho^{*} = 1.50 \ \text{and}\ a = 1.37 $} \\ \cline{2-5} \cline{6-9}
 {}  & & $\ \mid\vec{G}\mid$ &\quad ODS &\quad WCAS & & $\ \mid\vec{G}\mid$  &\quad ODS &\quad WCAS  \\ \hline
 $ 0 $ &  & $   0.000 $ & $  -52.40 $ & $  -48.24 $ &  & $  0.000 $ & $  -39.45 $ & $  -38.66 $\\
 $ 1 $ &  & $   7.021 $ & $   0.637 $ & $   0.632 $ &  & $  7.944 $ & $   0.435 $ & $   0.436 $\\
 $ 2 $ &  & $   8.107 $ & $   0.144 $ & $   0.160 $ &  & $  9.173 $ & $   0.102 $ & $   0.104 $\\
 $ 3 $ &  & $  11.466 $ & $  -0.204 $ & $  -0.225 $ &  & $ 12.972 $ & $  -0.128 $ & $  -0.134 $\\
 $ 4 $ &  & $  13.445 $ & $   0.253 $ & $   0.267 $ &  & $ 15.211 $ & $   0.163 $ & $   0.170 $\\
 $ 5 $ &  & $  14.042 $ & $   0.168 $ & $   0.174 $ &  & $ 15.887 $ & $   0.108 $ & $   0.112 $\\
 $ 6 $ &  & $  16.215 $ & $  -0.204 $ & $  -0.221 $ &  & $ 18.345 $ & $  -0.123 $ & $  -0.129 $\\
 $ 7 $ &  & $  17.669 $ & $  -0.067 $ & $  -0.064 $ &  & $ 19.991 $ & $  -0.041 $ & $  -0.040 $\\
 $ 8 $ &  & $  18.129 $ & $   0.002 $ & $   0.010 $ &  & $ 20.510 $ & $   0.000 $ & $   0.002 $\\
 $ 9 $ &  & $  19.859 $ & $   0.106 $ & $   0.111 $ &  & $ 22.468 $ & $   0.063 $ & $   0.066 $\\
$ 10 $ &  & $  21.064 $ & $   0.006 $ & $  -0.001 $ &  & $ 23.831 $ & $   0.007 $ & $   0.004 $\\
$ 11 $ &  & $  22.931 $ & $  -0.077 $ & $  -0.082 $ &  & $ 25.944 $ & $  -0.044 $ & $  -0.045 $\\
$ 12 $ &  & $  23.982 $ & $  -0.029 $ & $  -0.026 $ &  & $ 27.133 $ & $  -0.019 $ & $  -0.018 $\\
$ 13 $ &  & $  24.322 $ & $  -0.008 $ & $  -0.002 $ &  & $ 27.518 $ & $  -0.007 $ & $  -0.005 $\\
$ 14 $ &  & $  25.638 $ & $   0.049 $ & $   0.055 $ &  & $ 29.006 $ & $   0.026 $ & $   0.028 $\\
$ 15 $ &  & $  26.582 $ & $   0.039 $ & $   0.039 $ &  & $ 30.074 $ & $   0.023 $ & $   0.023 $\\
\end{tabular}
\end{ruledtabular}
\end{table*}
\end{scriptsize}
\end{center}
\begin{center}
\LTcapwidth=\textwidth
\renewcommand{\arraystretch}{0.75}
\begin{longtable}{>{\scriptsize}c>{\scriptsize}c>{\scriptsize}c>{\scriptsize}c
>{\scriptsize}c>{\scriptsize}c>{\scriptsize}c}
\caption{Comparison of freezing parameters $ \rho_{l}^{*} $, $ \rho_{s}^{*} $,  $ \Delta \rho^{*} $,  Lindemann parameter $   L   $ and pressure $  P^{*} = P\sigma^{3}/\epsilon $ found from the present theory with simulations\cite{kuijper-JCP-1990-WCA, ahmed-PRE-2009-WCA} and with the RY-DFT and the MWDA\cite{kuijper-JCP-1990} for the RLJ potential at several values of $ T^{*} $.}\\[1mm]
\hline \hline
$ T^{*}  $  &  Simulation/Theory Group   & $ \rho_{l}^{*} $ & $ \rho_{s}^{*} $ &  $ \Delta \rho^{*} $ & $   L   $ & $  P\sigma^{3}/\epsilon $ \\[1mm]
\hline
\endfirsthead
\multicolumn{7}{c}%
{\tablename\ \thetable\ -- \scriptsize{\textit{Continued from previous page}}} \\[1mm]
\hline \hline
$ T^{*}  $  &   Simulation/Theory Group   & $ \rho_{l}^{*} $ & $ \rho_{s}^{*} $ &  $ \Delta \rho^{*} $ & $   L   $ & $  P\sigma^{3}/\epsilon $ \\[1mm]
\hline
\endhead
\hline \hline \multicolumn{7}{c}{\scriptsize {\textit{Continued on next page}}} \\
\endfoot
\hline \hline \multicolumn{7}{c}{\scriptsize {\textit{\textbf{NOTE-}{* indicates values obtained from
interpolation of the tabulated values.}}}} \\
\endlastfoot 
\label{RLJ-Comp}
$    0.80   $  &      Present result      &  $   0.930 $  &  $ 0.988 $  &  $ 0.062 $  &  $ 0.092  $  &  $    9.68 $ \\
   { }          &       RY-DFT              &  $   0.988 $  &  $ 1.058 $  &  $ 0.070 $  &  $ 0.077  $  &  $   12.48 $ \\
   { }           &    MC Simulation \cite{ahmed-PRE-2009-WCA}     &  $   0.920 $  &  $ 0.990 $  &  $ 0.076 $  &  $   { }      $  &  $    9.60 $ \\
   { }           &    MC Simulation* \cite{kuijper-JCP-1990-WCA}     &  $   0.935 $  &  $ 1.009 $  &  $ 0.079 $  &  $   { }      $  &  $   10.27 $ \\ \hline
$    1.00   $  &      Present result      &  $   0.957 $  &  $ 1.016 $  &  $ 0.061 $  &  $ 0.093  $  &  $   12.52 $ \\
    { }          &       RY-DFT              &  $   1.022 $  &  $ 1.093 $  &  $ 0.069 $  &  $ 0.076  $  &  $   16.43 $ \\
    { }          &     MWDA Theory  \cite{kuijper-JCP-1990}     &  $   0.905 $  &  $ 1.015 $  &  $ 0.120 $  &  $ 0.103  $  &  $   10.40 $ \\
    { }          &    MC Simulation \cite{ahmed-PRE-2009-WCA}     &  $   0.950 $  &  $ 1.016 $  &  $ 0.069 $  &  $  { }       $  &  $   12.57 $  \\
    { }          &    MC Simulation \cite{kuijper-JCP-1990-WCA}     &  $   0.952 $  &  $ 1.023 $  &  $ 0.075 $  &  $  { }       $  &  $   12.60 $ \\ \hline
$    1.35   $  &      Present result      &  $   1.002 $  &  $ 1.059 $  &  $ 0.056 $  &  $ 0.094  $  &  $   18.07 $  \\
    { }          &        RY-DFT             &  $   1.076 $  &  $ 1.147 $  &  $ 0.066 $  &  $ 0.075  $  &  $   24.17 $ \\
    { }          &    MC Simulation* \cite{ahmed-PRE-2009-WCA}     &  $   1.016 $  &  $ 1.086 $  &  $ 0.069 $  &  $  { }       $  &  $   19.46 $ \\      
    { }          &    MC Simulation* \cite{kuijper-JCP-1990-WCA}     &  $   0.988 $  &  $ 1.056 $  &  $ 0.069 $  &  $  { }       $  &  $   17.70 $ \\ \hline
$    1.50   $  &      Present result      &  $   1.019 $  &  $ 1.075 $  &  $ 0.055 $  &  $ 0.095  $  &  $   20.52 $ \\
    { }          &         RY-DFT            &  $   1.095 $  &  $ 1.167 $  &  $ 0.066 $  &  $ 0.076  $  &  $   27.57 $ \\
    { }          &    MC Simulation* \cite{ahmed-PRE-2009-WCA}     &  $   1.034 $  &  $ 1.104 $  &  $ 0.067 $  &  $    { }     $  &  $   22.11 $ \\
    { }          &    MC Simulation \cite{kuijper-JCP-1990-WCA}     &  $   1.010 $  &  $ 1.080 $  &  $ 0.069 $  &  $    { }     $  &  $   20.60 $ \\ \hline
$    2.00   $  &       Present result     &  $   1.066 $  &  $ 1.125 $  &  $ 0.055 $  &  $ 0.096  $  &  $   28.95 $  \\
    { }          &     RY-DFT     &  $   1.160 $  &  $ 1.231 $  &  $ 0.061 $  &  $ 0.077  $  &  $   40.83 $ \\
    { }          &     MWDA Theory  \cite{kuijper-JCP-1990}     &  $   1.050 $  &  $ 1.130 $  &  $ 0.080 $  &  $ 0.110  $  &  $   27.30  $ \\
    { }          &    MC Simulation \cite{ahmed-PRE-2009-WCA}     &  $   1.070 $  &  $ 1.140 $  &  $ 0.065 $  &  $   { }      $  &  $   30.40 $  \\
    { }          &    MC Simulation \cite{kuijper-JCP-1990-WCA}     &  $   1.087 $  &  $ 1.159 $  &  $ 0.066 $  &  $   { }      $  &  $   32.30 $ \\ \hline
$    2.74   $  &      Present result      &  $    1.128 $  &  $ 1.187 $  &  $ 0.052 $  &  $ 0.098  $  &  $   42.91 $  \\
    { }          &     RY-DFT      &  $    1.236 $  &  $ 1.311 $  &  $ 0.060 $  &  $ 0.077  $  &  $   62.23 $ \\
    { }          &    MC Simulation \cite{ahmed-PRE-2009-WCA}     &  $    1.130 $  &  $ 1.200 $  &  $ 0.062 $  &  $  { }       $  &  $   45.10 $  \\
    { }          &    MC Simulation* \cite{kuijper-JCP-1990-WCA}     &  $    1.176 $  &  $ 1.248 $  &  $ 0.061 $  &  $  { }       $  &  $   50.41 $ \\ \hline
$    4.00   $  &      Present result      &  $    1.214 $  &  $ 1.274 $  &  $ 0.049 $  &  $ 0.100  $  &  $   69.40 $  \\
    { }          &     RY-DFT     &  $    1.346 $  &  $ 1.421 $  &  $ 0.056 $  &  $ 0.078  $  &  $  105.14 $ \\
    { }          &    MC Simulation* \cite{ahmed-PRE-2009-WCA}     &  $    1.192 $  &  $ 1.245 $  &  $ 0.045 $  &  $   { }      $  &  $   75.53 $ \\
    { }          &    MC Simulation* \cite{kuijper-JCP-1990-WCA}     &  $    1.264 $  &  $ 1.333 $  &  $ 0.055 $  &  $   { }      $  &  $   80.71 $ \\ \hline
$    5.00   $  &       Present result     &  $    1.271 $  &  $ 1.331 $  &  $ 0.047 $  &  $ 0.101  $  &  $   92.49 $ \\
    { }          &      RY-DFT      &  $    1.416 $  &  $ 1.494 $  &  $ 0.055 $  &  $ 0.077  $  &  $  142.8 $ \\
    { }          &     MWDA Theory  \cite{kuijper-JCP-1990}     &  $    1.275 $  &  $ 1.350 $  &  $ 0.060 $  &  $ 0.110  $  &  $   93.90 $ \\
    { }          &    MC Simulation* \cite{ahmed-PRE-2009-WCA}     &  $    1.260 $  &  $ 1.317 $  &  $ 0.045 $  &  $     { }    $  &  $  102.1 $ \\
    { }          &    MC Simulation \cite{kuijper-JCP-1990-WCA}     &  $    1.304 $  &  $ 1.370 $  &  $ 0.051 $  &  $    { }     $  &  $  104.5 $ \\ \hline
$   10.00   $  &      Present result      &  $    1.478 $  &  $ 1.539 $  &  $ 0.041 $  &  $ 0.107  $  &  $  227.8 $  \\
   { }           &       RY-DFT       &  $    1.671 $  &  $ 1.758 $  &  $ 0.052 $  &  $ 0.077  $  &  $  369.7 $  \\
   { }           &    MC Simulation* \cite{ahmed-PRE-2009-WCA}     &  $    1.495 $  &  $ 1.556 $  &  $ 0.041 $  &  $     { }    $  &  $  257.1 $ \\
\end{longtable}
\end{center}
\begin{center}
\LTcapwidth=\textwidth
\renewcommand{\arraystretch}{0.7}
\begin{longtable}{>{\scriptsize}c>{\scriptsize}c>{\scriptsize}c>{\scriptsize}c>{\scriptsize}
c>{\scriptsize}c>{\scriptsize}c}

\caption{Comparison of freezing parameters $ \rho_{l}^{*} $, $ \rho_{s}^{*} $,  $ \Delta \rho^{*} $,  Lindemann parameter $   L   $ and pressure $  P^{*} = P\sigma^{3}/\epsilon $ found from the present theory with simulations\cite{agrawal-MOLPHYS-1988, ahmed-JCP-2009, sousa-JCP-2012, hansen-PRev-1969, *hansen-PRA-1970} and with the RY-DFT and the MWDA\cite{kuijper-JCP-1990} for the LJ potential at different values of $ T^{*} $.}\\[1mm]

\hline \hline

$ T^{*}  $  &   Simulation/Theory Group   & $ \rho_{l}^{*} $ & $ \rho_{s}^{*} $ &  $ \Delta \rho^{*} $ & $   L   $ & $  P\sigma^{3}/\epsilon $ \\[1mm]
\hline
\endfirsthead
\multicolumn{7}{c}%
{\tablename\ \thetable\ -- \scriptsize{\textit{Continued from previous page}}} \\[1mm]
\hline \hline
$ T^{*}  $  &   Simulation/Theory Group   & $ \rho_{l}^{*} $ & $ \rho_{s}^{*} $ &  $ \Delta \rho^{*} $ & $   L   $ & $  P\sigma^{3}/\epsilon $ \\[1mm]
\hline
\endhead
\hline \hline \multicolumn{7}{c}{\scriptsize {\textit{Continued on next page}}} \\
\endfoot
\hline \hline \multicolumn{7}{c}{\scriptsize {\textit{\textbf{NOTE-}{* indicates values obtained from
interpolation of the tabulated values.}}}} \\
\endlastfoot 
\label{LJ-Comp}
   $ 0.80   $  &      Present result (ODS)    &  $   0.918 $  &  $ 0.976 $  &  $ 0.063 $  &  $ 0.091  $  &  $    2.57 $ \\
     { }    &      Present result (WCAS)   &  $   0.892 $  &  $ 0.960 $  &  $ 0.076 $  &  $ 0.091  $  &  $    1.65 $ \\
     { }    &           RY-DFT (ODS)       &  $   1.009 $  &  $ 1.073 $  &  $ 0.064 $  &  $ 0.074  $  &  $    6.39 $ \\
     { }    &           RY-DFT (WCAS)      &  $   0.963 $  &  $ 1.033 $  &  $ 0.073 $  &  $ 0.078  $  &  $    3.85 $ \\
     { }    &      MC Simulation* \cite{agrawal-MOLPHYS-1988}      &  $   0.878 $  &  $ 0.979 $  &  $ 0.115 $  &    { }      &  $    1.39 $ \\
     { }    &      MC Simulation \cite{ahmed-JCP-2009}      &  $   0.891 $  &  $ 0.983 $  &  $ 0.103 $  &    { }      &  $    1.65 $ \\
     { }    &      MC Simulation* \cite{sousa-JCP-2012}      &  $   0.875 $  &  $ 0.977 $  &  $ 0.117 $  &    { }      &  $    1.30 $ \\
     { }    &      MC Simulation* \cite{hansen-PRev-1969,*hansen-PRA-1970}      &  $   0.883 $  &  $ 0.979 $  &  $ 0.109 $  &    { }      &  $    1.23 $ \\ \hline
   $ 1.00  $  &      Present result (ODS)    &  $   0.957 $  &  $ 1.011 $  &  $ 0.056 $  &  $ 0.093  $  &  $    5.56 $ \\
     { }    &      Present result (WCAS)   &  $   0.929 $  &  $ 0.994 $  &  $ 0.069 $  &  $ 0.092  $  &  $    4.18 $ \\
     { }    &           RY-DFT (ODS)       &  $   1.053 $  &  $ 1.115 $  &  $ 0.059 $  &  $ 0.075  $  &  $   11.14 $ \\
     { }    &           RY-DFT (WCAS)      &  $   1.001 $  &  $ 1.073 $  &  $ 0.072 $  &  $ 0.077  $  &  $    7.33 $ \\
     { }    &    MWDA+MF Theory  \cite{kuijper-JCP-1990}      &  $   0.880 $  &  $ 1.025 $  &  $ 0.160 $  &  $ 0.100  $  &  $    3.20 $ \\
     { }    &      MC Simulation* \cite{agrawal-MOLPHYS-1988}      &  $   0.924 $  &  $ 1.010 $  &  $ 0.094 $  &    { }      &  $    4.11 $ \\
     { }    &      MC Simulation \cite{ahmed-JCP-2009}       &  $   0.923 $  &  $ 1.008 $  &  $ 0.092 $  &    { }      &  $    4.05 $ \\
     { }    &      MC Simulation \cite{sousa-JCP-2012}      &  $   0.920 $  &  $ 1.007 $  &  $ 0.095 $  &    { }      &  $    3.94 $ \\
     { }    &      MC Simulation* \cite{hansen-PRev-1969,*hansen-PRA-1970}      &  $   0.914 $  &  $ 1.004 $  &  $ 0.099 $  &    { }      &  $    3.59 $ \\ \hline
  $  1.35  $  &      Present result (ODS)    &  $   1.010 $  &  $ 1.062 $  &  $ 0.052 $  &  $ 0.094  $  &  $   11.20 $ \\
     { }    &      Present result (WCAS)   &  $   0.982 $  &  $ 1.045 $  &  $ 0.064 $  &  $ 0.094  $  &  $    9.11 $ \\
     { }    &           RY-DFT (ODS)       &  $   1.113 $  &  $ 1.178 $  &  $ 0.058 $  &  $ 0.074  $  &  $   20.03 $ \\      
     { }    &           RY-DFT (WCAS)      &  $   1.063 $  &  $ 1.134 $  &  $ 0.067 $  &  $ 0.077  $  &  $   14.57 $ \\
     { }    &      MC Simulation* \cite{agrawal-MOLPHYS-1988}      &  $   0.983 $  &  $ 1.061 $  &  $ 0.079 $  &    { }      &  $    9.53 $ \\
     { }    &      MC Simulation* \cite{ahmed-JCP-2009}       &  $   0.973 $  &  $ 1.049 $  &  $ 0.078 $  &    { }      &  $    8.99 $ \\
     { }    &      MC Simulation* \cite{sousa-JCP-2012}      &  $   0.981 $  &  $ 1.057 $  &  $ 0.077 $  &    { }      &  $    9.25 $ \\
     { }    &      MC Simulation \cite{hansen-PRev-1969,*hansen-PRA-1970}      &  $   0.964 $  &  $ 1.053 $  &  $ 0.092 $  &    { }      &  $    9.00 $ \\ \hline
   $ 1.50  $  &      Present result (ODS)    &  $   1.028 $  &  $ 1.082 $  &  $ 0.052 $  &  $ 0.095  $  &  $   13.70 $ \\
     { }    &      Present result (WCAS)   &  $   1.001 $  &  $ 1.064 $  &  $ 0.063 $  &  $ 0.094  $  &  $   11.35 $ \\
     { }    &           RY-DFT (ODS)       &  $   1.137 $  &  $ 1.202 $  &  $ 0.057 $  &  $ 0.075  $  &  $   24.25 $ \\      
     { }    &           RY-DFT (WCAS)      &  $   1.085 $  &  $ 1.157 $  &  $ 0.066 $  &  $ 0.077  $  &  $   17.92 $ \\
     { }    &      MC Simulation* \cite{agrawal-MOLPHYS-1988}      &  $   1.007 $  &  $ 1.081 $  &  $ 0.074 $  &    { }      &  $   12.09 $ \\
     { }    &      MC Simulation \cite{ahmed-JCP-2009}       &  $   0.993 $  &  $ 1.069 $  &  $ 0.077 $  &    { }      &  $   11.20 $ \\
     { }    &      MC Simulation \cite{sousa-JCP-2012}      &  $   1.002 $  &  $ 1.076 $  &  $ 0.074 $  &    { }      &  $   11.75 $ \\
     { }    &      MC Simulation* \cite{hansen-PRev-1969,*hansen-PRA-1970}      &  $   0.984 $  &  $ 1.073 $  &  $ 0.091 $  &    { }      &  $   11.52 $ \\ \hline
   $ 2.00  $  &      Present result (ODS)    &  $   1.084 $  &  $ 1.139 $  &  $ 0.052 $  &  $ 0.096  $  &  $   22.84 $ \\
     { }    &      Present result (WCAS)   &  $   1.060 $  &  $ 1.120 $  &  $ 0.057 $  &  $ 0.096  $  &  $   19.76 $ \\
     { }    &           RY-DFT (ODS)       &  $   1.207 $  &  $ 1.271 $  &  $ 0.053 $  &  $ 0.075  $  &  $   39.60 $ \\      
     { }    &           RY-DFT (WCAS)      &  $   1.155 $  &  $ 1.225 $  &  $ 0.061 $  &  $ 0.077  $  &  $   30.54 $ \\
     { }    &    MWDA+MF Theory  \cite{kuijper-JCP-1990}      &  $   1.040 $  &  $ 1.140 $  &  $ 0.100 $  &  $ 0.111  $  &  $   17.70 $ \\
     { }    &      MC Simulation* \cite{agrawal-MOLPHYS-1988}      &  $   1.071 $  &  $ 1.139 $  &  $ 0.064 $  &    { }      &  $   21.35 $ \\
     { }    &      MC Simulation* \cite{ahmed-JCP-2009}       &  $   1.050 $  &  $ 1.124 $  &  $ 0.071 $  &    { }      &  $   19.45 $ \\
     { }    &      MC Simulation \cite{sousa-JCP-2012}      &  $   1.065 $  &  $ 1.134 $  &  $ 0.065 $  &    { }      &  $   20.81 $ \\
     { }    &      MC Simulation* \cite{hansen-PRev-1969,*hansen-PRA-1970}      &  $   1.042 $  &  $ 1.125 $  &  $ 0.079 $  &    { }      &  $   19.59 $ \\ \hline
  $  2.74  $  &      Present result (ODS)    &  $   1.152 $  &  $ 1.207 $  &  $ 0.047 $  &  $ 0.098  $  &  $   37.86 $ \\
     { }    &      Present result (WCAS)   &  $   1.126 $  &  $ 1.188 $  &  $ 0.055 $  &  $ 0.098  $  &  $   33.25 $ \\
     { }    &           RY-DFT (ODS)       &  $   1.288 $  &  $ 1.356 $  &  $ 0.053 $  &  $ 0.075  $  &  $   64.31 $ \\      
     { }    &           RY-DFT (WCAS)      &  $   1.235 $  &  $ 1.307 $  &  $ 0.058 $  &  $ 0.077  $  &  $   51.18 $ \\
     { }    &      MC Simulation \cite{agrawal-MOLPHYS-1988}      &  $   1.144 $  &  $ 1.211 $  &  $ 0.059 $  &    { }      &  $   36.91 $ \\
     { }    &      MC Simulation \cite{ahmed-JCP-2009}       &  $   1.116 $  &  $ 1.181 $  &  $ 0.058 $  &    { }      &  $   33.20 $ \\
     { }    &      MC Simulation* \cite{sousa-JCP-2012}      &  $   1.139 $  &  $ 1.206 $  &  $ 0.059 $  &    { }      &  $   35.95 $ \\
     { }    &      MC Simulation \cite{hansen-PRev-1969,*hansen-PRA-1970}      &  $   1.113 $  &  $ 1.179 $  &  $ 0.059 $  &    { }      &  $   32.20 $ \\ \hline
   $ 4.00  $  &      Present result (ODS)    &  $   1.241 $  &  $ 1.297 $  &  $ 0.045 $  &  $ 0.101  $  &  $   65.82 $ \\
     { }    &      Present result (WCAS)   &  $   1.216 $  &  $ 1.278 $  &  $ 0.051 $  &  $ 0.100  $  &  $   59.19 $ \\
     { }    &           RY-DFT (ODS)       &  $   1.400 $  &  $ 1.471 $  &  $ 0.051 $  &  $ 0.074  $  &  $  113.1 $ \\     
     { }    &           RY-DFT (WCAS)      &  $   1.346 $  &  $ 1.424 $  &  $ 0.058 $  &  $ 0.076  $  &  $   92.84 $ \\
     { }    &      MC Simulation* \cite{agrawal-MOLPHYS-1988}      &  $   1.245 $  &  $ 1.309 $  &  $ 0.052 $  &    { }      &  $   66.78 $ \\
     { }    &      MC Simulation \cite{sousa-JCP-2012}      &  $   1.237 $  &  $ 1.303 $  &  $ 0.053 $  &    { }      &  $   65.37 $ \\
     { }    &      MC Simulation* \cite{hansen-PRev-1969,*hansen-PRA-1970}      &  $   1.213 $  &  $ 1.274 $  &  $ 0.050 $  &    { }      &  $   59.93 $ \\ \hline
  $  5.00  $  &      Present result (ODS)    &  $   1.297 $  &  $ 1.354 $  &  $ 0.044 $  &  $ 0.102  $  &  $   89.53 $ \\
     { }    &      Present result (WCAS)   &  $   1.275 $  &  $ 1.336 $  &  $ 0.048 $  &  $ 0.102  $  &  $   82.04 $ \\
     { }    &           RY-DFT (ODS)       &  $   1.471 $  &  $ 1.546 $  &  $ 0.051 $  &  $ 0.074  $  &  $  155.5 $ \\      
     { }    &           RY-DFT (WCAS)      &  $   1.420 $  &  $ 1.495 $  &  $ 0.053 $  &  $ 0.077  $  &  $  130.6 $ \\
     { }    &    MWDA+MF Theory  \cite{kuijper-JCP-1990}      &  $   1.270 $  &  $ 1.350 $  &  $ 0.060 $  &  $ 0.110  $  &  $   79.80 $ \\
     { }    &      MC Simulation* \cite{agrawal-MOLPHYS-1988}      &  $   1.306 $  &  $ 1.373 $  &  $ 0.051 $  &    { }      &  $   93.13 $ \\
     { }    &      MC Simulation \cite{sousa-JCP-2012}      &  $   1.300 $  &  $ 1.366 $  &  $ 0.051 $  &    { }      &  $   91.20 $ \\
     { }    &      MC Simulation \cite{hansen-PRev-1969,*hansen-PRA-1970}      &  $   1.279 $  &  $ 1.349 $  &  $ 0.055 $  &    { }      &  $   86.00 $ \\ \hline
 $  10.0  $  &      Present result (ODS)    &  $   1.506 $  &  $ 1.562 $  &  $ 0.037 $  &  $ 0.107  $  &  $  231.3 $ \\
     { }    &      Present result (WCAS)   &  $   1.485 $  &  $ 1.547 $  &  $ 0.042 $  &  $ 0.107  $  &  $  215.9 $ \\
     { }    &           RY-DFT (ODS)       &  $   1.729 $  &  $ 1.812 $  &  $ 0.048 $  &  $ 0.074  $  &  $  409.6 $ \\      
     { }    &           RY-DFT (WCAS)      &  $   1.668 $  &  $ 1.756 $  &  $ 0.053 $  &  $ 0.077  $  &  $  347.3 $ \\
     { }    &    MWDA+MF Theory  \cite{kuijper-JCP-1990}      &  $   1.530 $  &  $ 1.580 $  &  $ 0.040 $  &  $ 0.104  $  &  $  242.0 $ \\
     { }    &       MWDA Theory  \cite{kuijper-JCP-1990}      &  $   1.520 $  &  $ 1.570 $  &  $ 0.030 $  &  $ 0.114  $  &  $  237.0 $ \\
     { }    &      MC Simulation* \cite{agrawal-MOLPHYS-1988}      &  $   1.530 $  &  $ 1.599 $  &  $ 0.045 $  &    { }      &  $  248.0 $ \\
     { }    &      MC Simulation \cite{hansen-PRev-1969,*hansen-PRA-1970}      &  $   1.500 $  &  $ 1.572 $  &  $ 0.048 $  &    { }      &  $  231.0 $ \\
\end{longtable}
\end{center}
\begin{center}
\begin{scriptsize}
\begin{table}
\caption{Comparison of values of  $ \frac{W_{id}}{N} $,  $ \frac{W_{0}}{N} $, and $ \frac{W_{b}}{N} $ at the freezing point for the RLJ potential and for the LJ potential(found from WCAS).}
\vspace{0.5cm}
\label{RLJ-Data}
\begin{ruledtabular}
\begin{tabular}{>{\scriptsize}c>{\scriptsize}c>{\scriptsize}c>{\scriptsize}c>{\scriptsize}c>{\scriptsize}c>{\scriptsize}c>{\scriptsize}c>{\scriptsize}c>{\scriptsize}c>{\scriptsize}c}
\small
 
 {   } & & \mc{4}{c}{ RLJ Potential} & & \mc{4}{c}{ LJ Potential}  \\ \cline{2-6} \cline{7-11}
$ T^{*} $ & & $ \rho_{s}^{*} $ &   $ \frac{W_{id}}{N} $  & $ \frac{W_{0}}{N} $ & $ \frac{W_{b}}{N} $ & & $ \rho_{s}^{*} $ &   $ \frac{W_{id}}{N} $  & $ \frac{W_{0}}{N} $ & $ \frac{W_{b}}{N} $  \\
\hline

 $  0.80 $  &  &  $ 0.988 $  &  $ 4.471 $  &  $ -3.972 $  &  $ -0.499 $ & &  $ 0.960 $  &  $ 4.562 $  &  $ -4.115 $  &  $ -0.449 $ \\
 $  1.00 $  &  &  $ 1.016 $  &  $ 4.437 $  &  $ -3.932 $  &  $ -0.506 $ & &  $ 0.994 $  &  $ 4.494 $  &  $ -4.042 $  &  $ -0.454 $ \\
 $  1.35 $  &  &  $ 1.059 $  &  $ 4.372 $  &  $ -3.860 $  &  $ -0.513 $ & &  $ 1.045 $  &  $ 4.419 $  &  $ -3.964 $  &  $ -0.458 $ \\
 $  1.50 $  &  &  $ 1.075 $  &  $ 4.351 $  &  $ -3.839 $  &  $ -0.515 $ & &  $ 1.064 $  &  $ 4.396 $  &  $ -3.938 $  &  $ -0.460 $ \\
 $  2.00 $  &  &  $ 1.125 $  &  $ 4.303 $  &  $ -3.785 $  &  $ -0.520 $ & &  $ 1.120 $  &  $ 4.316 $  &  $ -3.850 $  &  $ -0.468 $ \\
 $  2.74 $  &  &  $ 1.187 $  &  $ 4.235 $  &  $ -3.711 $  &  $ -0.525 $ & &  $ 1.188 $  &  $ 4.246 $  &  $ -3.768 $  &  $ -0.480 $ \\
 $  4.00 $  &  &  $ 1.274 $  &  $ 4.149 $  &  $ -3.620 $  &  $ -0.530 $ & &  $ 1.278 $  &  $ 4.153 $  &  $ -3.660 $  &  $ -0.494 $ \\
 $  5.00 $  &  &  $ 1.331 $  &  $ 4.095 $  &  $ -3.563 $  &  $ -0.533 $ & &  $ 1.336 $  &  $ 4.094 $  &  $ -3.593 $  &  $ -0.502 $ \\
 $ 10.00 $  &  &  $ 1.539 $  &  $ 3.919 $  &  $ -3.383 $  &  $ -0.539 $ & &  $ 1.547 $  &  $ 3.918 $  &  $ -3.399 $  &  $ -0.520 $ \\
\end{tabular}
\end{ruledtabular}
\end{table}
\end{scriptsize}
\end{center}

\begin{figure}[ht]
\includegraphics[height=3.0in, width=4.0in,clip]{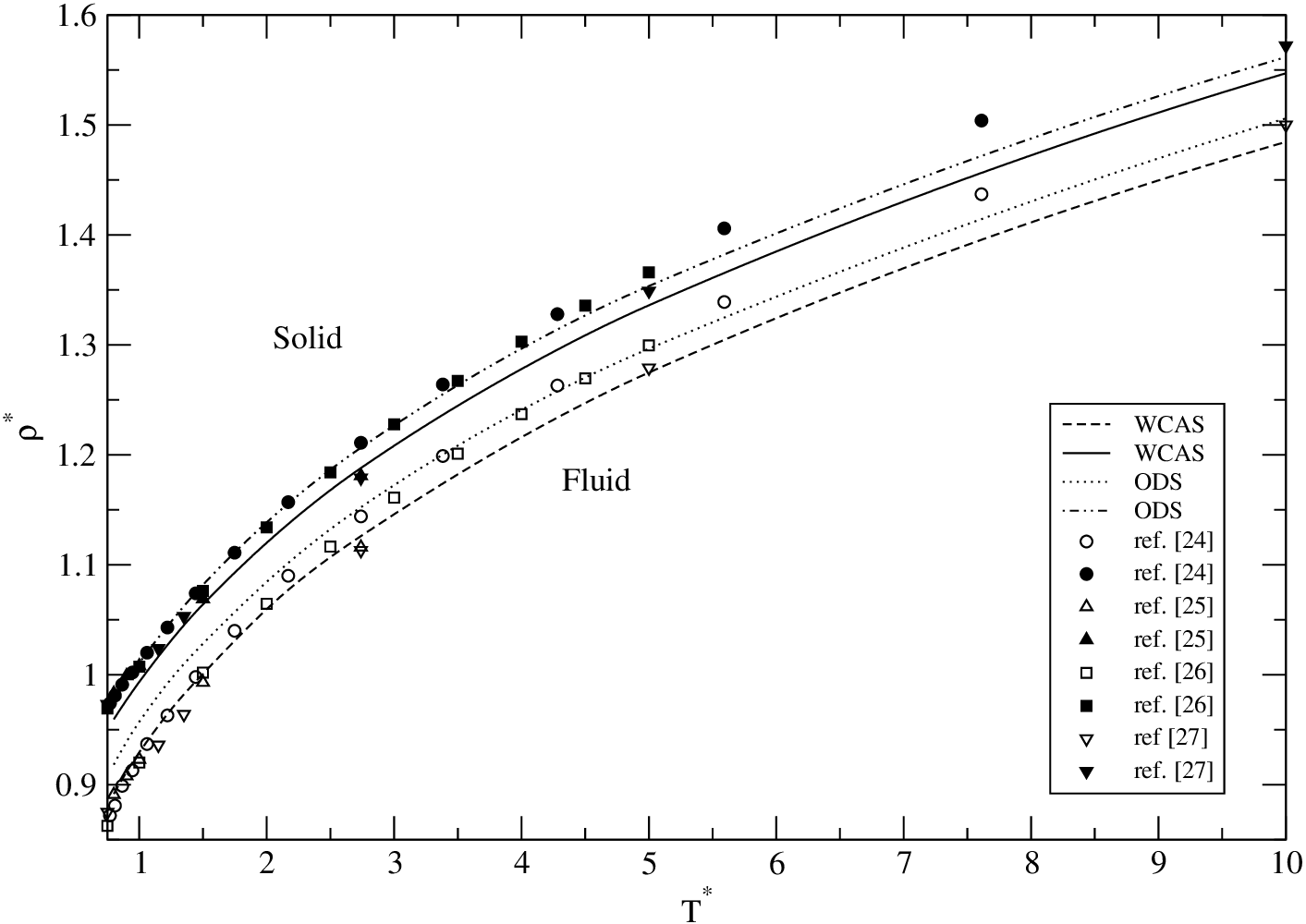}
\caption{The $ T^{*}-\rho^{*} $ phase diagram for the LJ potential in which lines represent present data (calculated via WCAS and ODS) and symbols represent simulation data (full for crystal and open for fluid) found from\citet{agrawal-MOLPHYS-1988}, \citet{ahmed-JCP-2009}, \citet{sousa-JCP-2012} and Hansen \cite{hansen-PRev-1969,*hansen-PRA-1970}.}
\label{fig-9}
\end{figure}

\begin{figure}[ht]
\includegraphics[height=3.0in, width=4.0in,clip]{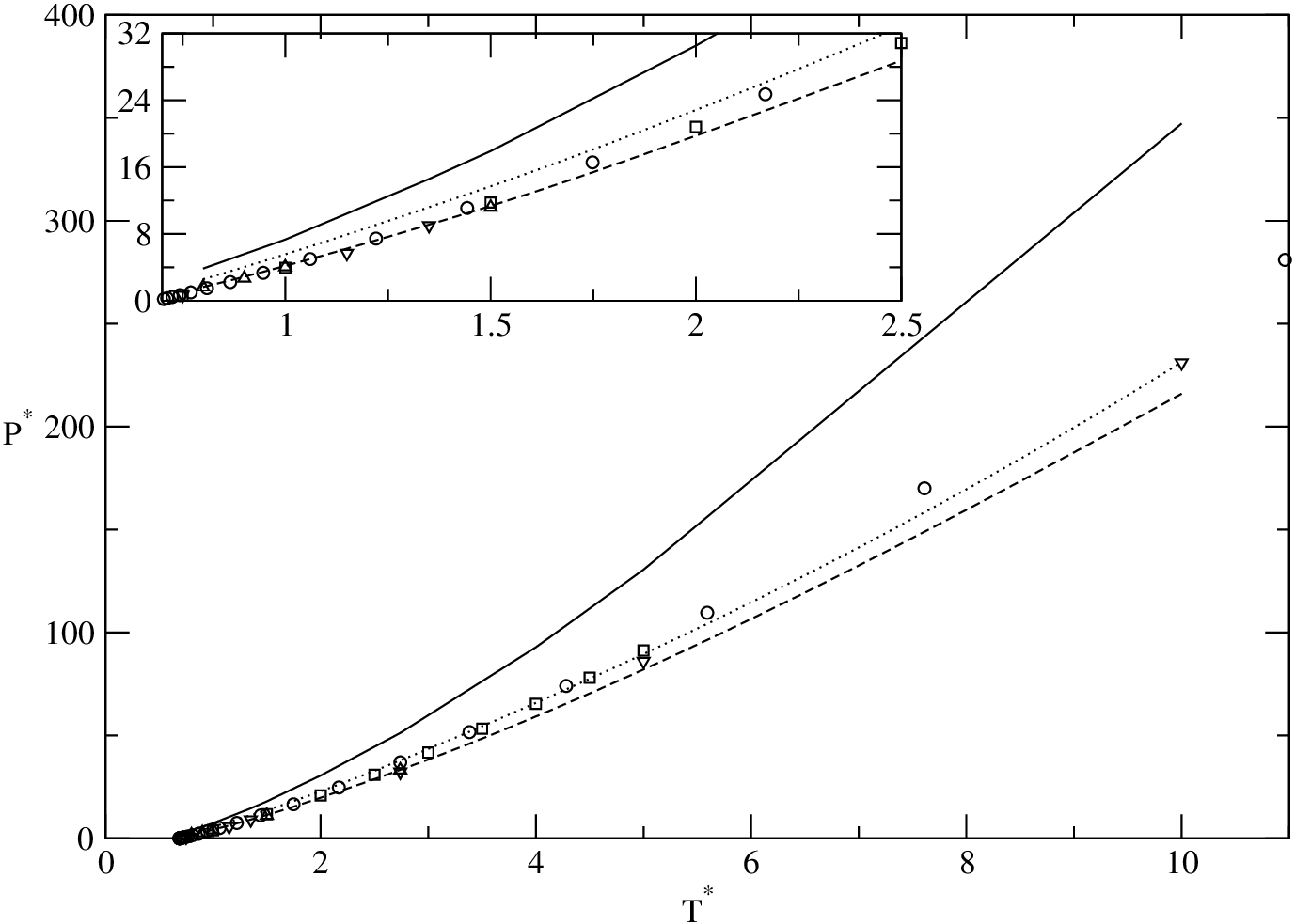}
\caption{Pressure $P^{*}$ \textit{vs} temperature $ T^{*} $  for LJ potential. Dashed and dotted lines represent present data found from WCAS and ODS division of LJ potential respectively and full line represents data found from RY-DFT. Symbols represent simulation data found from \cite{agrawal-MOLPHYS-1988,ahmed-JCP-2009,sousa-JCP-2012,hansen-PRev-1969,*hansen-PRA-1970}; notations are same as in Fig. \ref{fig-9}.}
\label{fig-10}
\end{figure}

\end{document}